# High-Resolution Non-Invasive X-ray Diffraction Analysis of Artists' Paints

Craig Hiley[a], Graeme Hansford[a], Nicholas Eastaugh[b]

[a]University of Leicester, Space Research Centre, School of Physics and Astronomy, University Road, Leicester LE1 7RH, UK.
[b]Art Discovery Inc., 162-164 Abbey Street, Bermondsey, London SE1 2AN, UK.

Corresponding author: Graeme Hansford (gmh14@leicester.ac.uk)

Abstract

Energy-dispersive X-ray diffraction (EDXRD) is extremely insensitive to sample morphology when implemented in a back-reflection geometry. The capabilities of this non-invasive technique for cultural heritage applications have been explored at high resolution at the Diamond Light Source synchrotron. The results of the XRD analysis of the pigments in 40 paints, commonly used by 20$^{th}$ century artists, are reported here. It was found that synthetic organic pigments yielded weak diffraction patterns at best, and it was not possible to unambiguously identify any of these pigments. In contrast, the majority of the paints containing inorganic pigments yielded good diffraction patterns amenable to crystallographic analysis. The high resolution of the technique enables the extraction of a range of detailed information: phase identification (including solid solutions), highly accurate unit cell parameters, phase quantification, crystallite size and strain parameters and preferred orientation parameters. The implications of these results for application to real paintings are discussed, along with the possibility to transfer the technique away from the synchrotron and into the laboratory and museum through the use of state-of-the-art microcalorimeter detectors. The results presented demonstrate the exciting potential of the technique for art history and authentication studies, based on the non-invasive acquisition of very high quality crystallographic data.

## Research Aim

The back-reflection EDXRD technique offers a method of retrieving highly accurate crystallographic parameters irrespective of the sample morphology. The method is suited to the non-invasive study of archaeological artefacts and artworks, especially paintings. We have conducted a survey study of forty oil-based artists' paints on a test panel, as a forerunner to the investigation of paintings, to explore and understand the advantages and limitations of this technique in the context of the art history and authentication of paintings.

## 1. Introduction

Scientific analysis of artworks can aid in their authentication [1] and attribution [2], inform conservation and restoration strategies [3] and yield insights into historical manufacturing practices [4] and painters' techniques [5]. In particular, analysis of the chemical composition of paints and pigments has proven a rich source of information. Pigments are an incredibly diverse family of materials ranging from organic dyes to minerals to synthetic inorganic salts [6]. A wide range of analytical techniques have been utilised in the scientific analysis of paintings, including hyperspectral imaging, X-ray fluorescence (XRF) spectroscopy, Fourier transform infrared (FTIR) spectroscopy and Raman spectroscopy [7, 8]. These techniques can be employed non-invasively and each has its own advantages and disadvantages. For example, XRF methods yield elemental information that in some instances allows the presence of specific pigments to be inferred but does not yield any further insights on structural characteristics. Raman methods are widely used and can identify specific pigments, notably synthetic organic pigments [9], but is ineffective for some inorganic pigments and does not generally allow further analysis such as quantification [10]. In contrast, X-ray diffraction (XRD) is a powerful technique in materials science, allowing highly specific identification and providing additional structural information on crystalline materials, such crystallite morphology, size/strain parameters and phase weight ratios. XRD is ideally carried out on prepared, finely powdered samples in order to ensure that relative peak intensities are reproducible which aids phase identification and permits quantitative analysis. This constraint poses a challenge in the analysis of cultural heritage objects where there is a strong reluctance to alter the object by removing samples. If curators allow the extraction of samples from the artwork at all, it is usually limited to minute samples; such micro-sampling usually leaves no changes visible to the naked eye. However, sampling from a limited number of locations may not yield representative information about the entire artefact [4] and, in any case, completely non-invasive methods are the gold standard in the scientific examination of cultural heritage artefacts.

A variety of synchrotron-based XRD methods are applied to the non-invasive study of heritage artefacts [7, 11, 12] and, in addition, several groups have developed laboratory or portable XRD instrumentation for non-invasive analysis [13-20]. Conventional laboratory instruments are not in general designed to accommodate the variety of artefact shapes and sizes encountered, though on occasions the artefacts have convenient form factors [21]. Portable instruments have in common relatively low to moderate resolution of diffraction peaks when compared to commercial laboratory diffractometers. The most common design [19, 22] applies angle-dispersive XRD with a shallow incidence angle of the X-ray beam on the artefact. The instrument must be positioned at a precise distance from the surface being studied, typically achieved using laser pointing devices. Uneven surfaces, such as impasto on paintings, can present challenges to this type of instrument and adversely affects resolution and precision. The modest resolution achievable commonly limits portable XRD to phase identification [13,14,18,19] though in some instances depth information can be retrieved [15,16] as well as texture and crystallinity [16].

Energy-dispersive XRD (EDXRD) in a back-reflection geometry has previously been shown to have very low dependence on the morphology of the sample [23-25], a characteristic that is highly desirable in the context of archaeometry. Furthermore, very high resolution can be achieved, while retaining insensitivity to sample morphology, by implementing the technique at a synchrotron facility [25]. The need to transport paintings and other irreplaceable heritage objects to a synchrotron is a disadvantage and in many cases is not allowed by galleries and museums or involves high insurance costs. However, we plan ultimately to transfer the back-reflection EDXRD technique to the laboratory and museum though the use of microcalorimeter technology. The motivation for the synchrotron-based study presented here is to explore the advantages and limitations of the technique applied to paintings, initially using representative paint samples, and to support our plans to develop a transportable laboratory-based instrument capable of high resolution.

Throughout this article the term 'non-invasive' is intended to indicate that no material changes at all are made to the object under study. Although by no means universal, some authors have used 'non-destructive' to mean that the analytical technique in question does not *consume* the sample which is therefore subsequently available to other techniques, but not excluding the possibility that the sample has been prepared in some way prior to the measurement. For this reason, 'non-invasive' is the preferred term here.

Details of the experimental set-up and data processing are presented in Section 2 of this article. The results for six paints are highlighted in Section 3, illustrating the type of information that can be retrieved using the back-reflection EDXRD technique. The data for the remaining paint samples can be found in the Supplementary Information. In Section 4 we discuss the significance of these results and the complicating factors to be expected in the analysis of paintings as opposed to paint samples. We also highlight the prospects of transferring the technique into the laboratory and museum with comparable performance characteristics. Lastly, the conclusions of this study are presented in Section 5.

## 2. Materials and Methods

The data was acquired over two separate beamtime sessions on the B18 beamline [26] at the Diamond Light Source synchrotron in February 2019 and January 2020. The experimental configuration and the data preparation methods were based on the principles described in Hansford *et al* [25] but with some differences in the details, as described below.

### 2.1. Artists' Paints

A pre-existing test panel with 41 oil-based paints commonly used by 20[th] century artists provided a representative range of pigment types. A full description of the paints is given by Polak *et al* [8] and additional details are given in the Supplementary Information. Diffraction data was acquired for each paint on the panel except Barite White (in this article the suppliers' paint names have been capitalised) which showed clear signs of degradation. The paints have been applied generously with significant non-planar morphology of up to 5 mm relative to the panel surface. Summary information for each paint is shown in Table 1.

### 2.2. Experimental Configuration

The optimum implementation of the back-reflection EDXRD method requires an annular detector, with the incident beam passing through a central aperture, because this geometry simultaneously maximises both the $2\theta$ angle and the detector area. PNDetector's Rococo-2 four-channel silicon drift detector (SDD) assembly provides an excellent approximation to an annular design. The Rococo-2 has an active area of 60 mm$^2$ and a central aperture of 1.8 mm diameter. The detector assembly was mounted in a purpose-designed vacuum chamber that was incorporated onto the B18 beamline. To allow the greatest flexibility in the size, positioning and translation of samples, the experiment was designed so that objects of interest are mounted in free-space. The chamber therefore has a thin vacuum window mounted at the end of a small 'snout' and samples should ideally be placed within a few millimetres of the window in order to minimise X-ray absorption within the air gap. For many samples it was possible to achieve a 2 – 3 mm gap and 5 mm is considered the worst case in practice. For a 5 mm gap, the sample-detector distance is 155 mm, a considerable improvement on the previous configuration [25]. When combined with the larger detector area, the solid angle subtended by the detector has been increased six-fold. A photograph of the experimental configuration is shown in Fig. S2.

**Table 1.** Summary of the artists' paints including phases identified in this work.

| Paint Name | Supplier | Phases Identified | Group Assignment[a] |
|---|---|---|---|
| **February 2019** | | | |
| Aureolin | Michael Harding | $K_3[Co(NO_2)_6]$ | A |
| Cadmium Gold Yellow | Michael Harding | CdS (greenockite) | A |
| Cadmium Red | Michael Harding | $CdS_{1-x}Se_x$ | A |
| Cerulean Blue | Michael Harding | $Co(Cr_xAl_{1-x})_2O_4$ | A |
| | | $Al_2O_3$ (corundum) | |
| | | $BaSO_4$ (barite) | |
| Chrome Green | Rublev Colours | $PbCr_{1-x}S_xO_4$ | A |
| Chrome Yellow | Rublev Colours | $PbCrO_4$ (crocoite) | A |
| | | $BaSO_4$ (barite) | |
| Dioxazine Violet | BLOCKX | $CaCO_3$ (calcite) | B |
| Flake White | Michael Harding | $2PbCO_3 \cdot Pb(OH)_2$ (hydrocerussite) | A |
| | | $PbCO_3$ (cerussite) | |
| | | ZnO (zincite) | |
| Flemish White | Rublev Colours | $(PbO)_3PbSO_4 \cdot H_2O$ | A |
| Ivory Black | Michael Harding | $Ca_5(PO_4)_3OH$ (hydroxyapatite) | A |
| | | $CaCO_3$ (calcite) | |
| Magenta | Michael Harding | - | D |
| Manganese Violet | Michael Harding | $\alpha,\beta$- $NH_4MnP_2O_7$ | A |
| Minium | Rublev Colours | $Pb_3O_4$ (minium) | A |
| Phthalocyanine Blue Lake | Michael Harding | - | D |
| Raw Sienna | Michael Harding | $\alpha$-FeOOH (goethite) | A |
| | | $SiO_2$ (quartz) | |
| | | $TiO_2$ (rutile) | |
| Raw Umber | Michael Harding | $\alpha$-FeOOH (goethite) | A |
| | | $SiO_2$ (quartz) | |
| Scarlet Lake | Michael Harding | - | D |
| Terre Vert | Michael Harding | - | C |
| Ultramarine Blue | Michael Harding | $(Na,Ca)_8[(S,Cl,SO_4,OH)_2\|Al_6Si_6O_{24})]$ (lazurite) | A |
| Vermilion | Rublev Colours | HgS (cinnabar) | A |
| Viridian | BLOCKX | - | C |
| Zinc White | Michael Harding | ZnO (zincite) | A |
| **January 2020** | | | |
| Alizarin Crimson | Michael Harding | - | D |
| Bright Yellow Lake | Michael Harding | $TiO_2$ (rutile) | B |
| Chrome Oxide Green | Michael Harding | $Cr_2O_3$ (eskolaite) | A |
| Cobalt Blue | Michael Harding | $CoAl_2O_4$ (spinel group) | A |
| | | $Al_2O_3$ (corundum) | |
| Cobalt Turquoise | Michael Harding | $CoCr_2O_4$ (spinel group) | A |
| | | $Cr_2O_3$ (eskolaite) | |
| Cremnitz White | Michael Harding | $2PbCO_3 \cdot Pb(OH)_2$ (hydrocerussite) | A |
| | | $PbCO_3$ (cerussite) | |
| French Yellow Ochre | Michael Harding | $\alpha$-FeOOH (goethite) | A |
| | | $SiO_2$ (quartz) | |
| | | $TiO_2$ (rutile) | |
| Lamp Black | Michael Harding | - | D |
| Lemon Yellow | Michael Harding | $BaCrO_4$ | A |
| Manganese Blue | BLOCKX | ZnO (zincite) | B |
| Naples Yellow | Michael Harding | $Pb_2Sb_2O_7$ (bindheimite) | A |
| | | $Pb_2SbSnO_{6.5}$ | |
| | | $BaSO_4$ (barite) | |
| Naphthol Red | Michael Harding | - | D |
| Orange Molybdate | Rublev Colours | - | C |
| Phthalocyanine Green Lake | Michael Harding | - | D |
| Pyrrolo Vermilion | BLOCKX | $TiO_2$ (rutile) | B |
| | | $CaCO_3$ (calcite) | |
| Titanium White #3 | Michael Harding | $TiO_2$ (rutile) | A |
| Transparent Red Oxide | Michael Harding | $Fe_2O_3$ (hematite) | A |
| Yellow Lake | Michael Harding | - | D |

[a]See Section 3 for the group definitions.

The use of a multi-element SDD brings an important advantage relative to a single SDD. Equal diffraction peak intensities on all four channels of the Rococo-2 is a reliable indicator of good powder averaging. Conversely, samples with large crystallites exhibit poor powder averaging with essentially random peak intensities – peaks can be completely missing on one or more detectors but present on others. Intermediate behaviour is also observed with the same peaks seen on all channels but with variable intensity. It is not unusual for one phase in a sample to show good powder averaging while another exhibits poor averaging (see, for example, Fig. S7), and indeed this result can potentially provide a means to assign peaks to one or other of the phases. These behaviours can provide valuable insights into the sample and phase characteristics.

The scan range of nominal X-ray energies for the results presented here was 2.1 – 5.1 keV, corresponding to $d$ = 2.95 – 1.22 Å. For some of the paint samples in the earlier beamtime session, the scan range was reduced to 2.7 ($d$ = 2.30 Å) to 5.1 keV because of a loss of signal at lower energies, probably due to a minor air leak into a sealed section of the beamline. The time taken to complete each scan was about 20 mins. As in previous work [25], the beam was defocused to approximately 1.7 x 0.9 mm$^2$ in order to maximise powder averaging. The variation of the height of the sample surface within this area is likely to be below 1 mm for all the paints. The typical flux at the sample on B18 is ~5 x 10$^{11}$ ph s$^{-1}$ at 8 keV, with smaller values towards lower X-ray energies.

There was no visual indication of any radiation damage to the paints in these experiments. No additional tests were carried out to check for damage. Total radiation doses have been estimated using the equation given by Bertrand *et al* [27] (Table 1) and are around or below 5 MGy in all cases. Defocusing the beam is highly beneficial in reducing the potential for radiation damage. There have been only a small number of detailed studies of synchrotron radiation damage of pigments but the results have shown a damage threshold of ~10 MGy for chrome yellow [28] (worst case) and '10% level of damage' for ultramarine of 7.7 MGy [29]. Consequently, the risk of radiation damage in the experiments reported here is minimal.

2.2.1. February 2019 beamtime

For this beamtime, the vacuum window was a nominally 3 μm thick carbon foil provided by Nannotek (Micromatter Technologies Inc.). The calculated total attenuation of a diffracted X-ray beam for a double-pass through this window plus a 5 mm air gap is shown in Fig. 1. The foil introduces a broad, relatively intense peak centred at 3691 eV corresponding to $d$ = 1.680 Å while no other features attributable to the foil appear in the diffraction patterns. It is surmised that the foil has a graphitic structure with highly-oriented basal planes parallel to the foil surface. The observed feature corresponds to the (004) diffraction peak of graphite while no other allowed basal plane reflections occur within the scanned energy range [30].

Alignment of the Rococo-2 assembly with the synchrotron beam was a difficult process involving manual manipulation of the position of the detector head. Unfortunately, it was subsequently found that three of the four individual SDD elements had suffered varying degrees of damage, most likely caused by micro-cracking induced by mechanical stresses during manipulation [Andreas Liebel (PNDetector), personal communication, 2019]. The data reported here was recorded by the unaffected detector, though one other detector yielded data of sufficient quality to allow the degree of powder averaging in each diffraction pattern to be assessed.

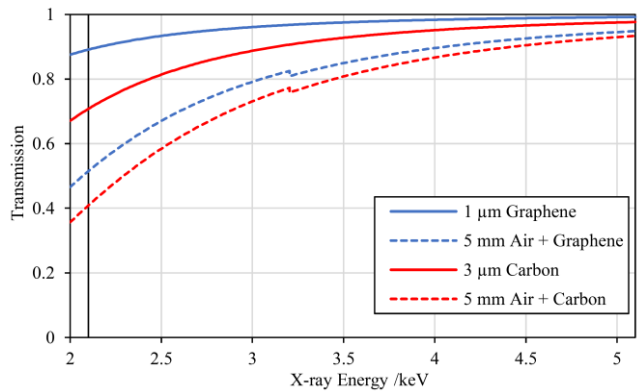

**Fig. 1.** Total X-ray attenuation for a double-pass through each X-ray window with and without a 5 mm air gap to the sample. The solid vertical line marks 2.1 keV, the lowest energy accessed in this work.

2.2.2. January 2020 beamtime

The damaged Rococo-2 assembly was replaced with an updated version incorporating xyz-translation stages, effectively solving the alignment difficulties and eliminating the risk of mechanical damage. The data from all four detector elements was of good quality, improving the overall signal-to-noise ratio and allowing good assessment of powder averaging, and the four channels were summed prior to analysis. Additionally, the 3 µm carbon foil window was replaced with a 1 µm graphene window supplied by Ketek GmbH [31]. The X-ray attenuation data for this window is shown in Fig. 1, illustrating that attenuation by the window itself now makes only a minor contribution to the total attenuation over the X-ray energies available on B18. Furthermore, the graphene window produces a much weaker and very broad graphitic (004) diffraction peak relative to the carbon foil window, centred at 3620 eV. In one further change, the 2.1 – 5.1 keV scan range was divided into two separate scans, joining at 2.8 keV, in order to bypass a Diamond data acquisition software limitation and increase the number of data points within each diffraction pattern and so avoid a larger-than-optimum step size across individual diffraction peaks.

**2.3. Data Preparation**

The preparation of the raw data generally followed the steps described previously [25]. TOPAS-Academic [32] was used for all crystallographic analysis of the diffraction patterns. The baseline-fitting capabilities of TOPAS were utilised instead of subtracting the baseline by fitting polynomial functions which can be a time-consuming process.

In order to obtain accurate *d*-spacings, it is vital to calibrate the nominal energy scale of the diffraction patterns. On B18, a Si double-crystal monochromator (DCM) was used to select a narrow X-ray energy bandpass utilising Si(111) diffraction [26]. The X-ray energy transmitted is dependent on the DCM angle according to the Bragg equation expressed in the energy domain:

$$E = \frac{hc}{2d_{Si(111)}\sin\theta_{DCM}} \qquad (1)$$

where $E$ is the X-ray energy, $h$ is the Planck constant, $c$ is the speed of light, $d_{Si(111)}$ is the *d*-spacing of the Si(111) reflection (taken to be 3.135693 Å [33]) and $\theta_{DCM}$ is the angle of the DCM relative to the incident beam. There can be a significant offset between the nominal DCM angle and the true angle which must be calibrated using diffraction standards.

2.3.1. February 2019 beamtime

A first, rough energy calibration was obtained by aligning the Ca K-edge XAFS spectrum from a marble sample with literature data for calcite [34]. In this step, the nominal DCM angle was shifted by -0.222° (the sign convention is defined below), to correct the energy scale to within ~1 eV. In addition, the experimental $2\theta$ angle was estimated to be 178°, based on the dimensions of the experimental components. A more accurate calibration was then carried out by Le Bail fits to the diffraction patterns of the National Institute of Standards and Technology (NIST) powder diffraction standard reference materials (SRMs) 640e (Si powder) [33] and 660c (LaB$_6$ powder) [35] in energy-space in TOPAS. The approximate DCM offset and $2\theta$ angles were used as a starting point for the refinements. Lattice parameters were fixed at the NIST-certified values, and the $\theta$ shift ($\delta\theta$) and DCM angle shift ($\delta\theta_{DCM}$) relative to the nominal values were optimised so that the calculated diffraction pattern best fitted the data, based on the relationship derived from the Bragg equation:

$$d_{SRM} = \frac{d_{Si(111)} \sin(\theta_{DCM} + \delta\theta_{DCM})}{\sin(\theta + \delta\theta)} \quad (2)$$

where $d_{SRM}$ is the $d$-spacing of each reflection of the diffraction standard. Fits to three EDXRD spectra (two acquired at the start of the beamtime and one towards the end) from the LaB$_6$ powder and two spectra from the Si powder were performed to obtain five values for $\delta\theta$ and $\delta\theta_{DCM}$ to ensure reproducibility (see Fig. S3). The refined values were averaged to give a final $\delta\theta_{DCM}$ of –0.2152 (2)° and $\delta(2\theta) = 2\delta\theta = $ –0.63 (3)°, yielding $2\theta = $ 177.37 (3)°. No statistically significant difference in the DCM or $2\theta$ angles was observed in spectra from the SRMs between the start and end of the experiment.

2.3.2. January 2020 beamtime

The same calibration procedure was followed for the more recent data except that there was no need for an initial rough calibration. Two EDXRD spectra of the LaB$_6$ powder (one acquired at the start and one towards the end) were fitted to yield $\delta\theta_{DCM} = $ –0.006853 (7)°. When $\delta\theta$ was included in the fits it was highly correlated with $\delta\theta_{DCM}$ and made virtually no difference to the weighted-profile reliability factor $R_{wp}$; it was therefore fixed at zero without compromise to the accuracy of the calibration.

**2.4. XRD Analysis**

For simplicity within TOPAS and easier comparison with conventional XRD data, diffraction patterns were converted from energy-space to pseudo-angular-dispersive datasets using a dummy wavelength of 1.5406 Å (corresponding to Cu K$\alpha_1$, 8.0478 keV). Subsequent references to the $2\theta$ angle alludes to the converted data rather than a physical angle. Where available, information from the paint supplier and data from previous characterisation of the paint samples (Raman spectroscopy and elemental analysis [8]) were used to determine the phases likely to be present. Literature crystal structures for these phases were obtained from the Inorganic Crystal Structure Database (ICSD) [36] and used as structure models in TOPAS fits to the data. All diffraction peaks were fitted using the modified Thompson-Cox-Hastings pseudo-Voigt (TCHZ) peak shape [37]. The broad (004) graphite diffraction peaks attributed to the vacuum windows were handled by fitting either one or two (January 2020 or February 2019 data respectively) pseudo-Voigt peaks independently of the sample phases.

2.4.1. Crystallite Size Analysis

Fits to the diffraction data of the NIST LaB$_6$ SRM and two ZnO powder SRMs with certified median crystalline domain diameters of 60 nm and 15 nm (SRM 1979 [38]) were used to validate the accuracy of the crystallite size analysis based on peak broadening. A Pawley fit to the LaB$_6$ diffraction pattern was used to determine the instrument broadening as a function of $2\theta$, using the formulation specified by Thompson-Cox-Hastings [37]. The instrument- and sample-broadening functions were then convoluted to analyse the ZnO data using a Pawley fit. The Scherrer equation allows the volume-weighted mean crystalline domain diameter, $L_{vol}$ (which can differ significantly from the median, depending on the size distribution [39]), to be obtained from the sample-broadening function, giving values of 109 (13) nm and 22 (2) nm for the '60 nm' and '15 nm' samples respectively. The $L_{vol}$ values reported in the NIST Certificate of Analysis [38], measured using two different diffractometers, are 138.9 (6) nm and 97.2 (14) nm for the '60 nm' sample and 31.39 (9) nm and 31.7 (5) nm for the '15 nm' sample. The level of consistency with the NIST values is considered sufficient for the purposes of this study. The crystalline domain diameters given in this paper are all volume-weighted except as noted below.

The crystallite shape model of Ectors *et al* [40] was fitted to a small number of phases with anisotropic peak broadening. In each of these cases the phase has trigonal symmetry (indexed using hexagonal coordinates) and a cylindrical crystallite shape was assumed with diameter $L_x$ and length $L_z$. The cylinder axis is parallel to the <001> direction. The Ectors model provides area-weighted crystallite dimensions.

2.4.2. Phase Fraction Analysis

It is well known that quantitative phase analysis using EDXRD data is challenging because the dependence of the relevant physical quantities on X-ray energy are often not accurately known and may be difficult to calibrate [41]. The paint samples typically consist of crystalline pigments and fillers in an amorphous dried oil matrix in unknown relative amounts and the absorption coefficient cannot therefore be calculated *a priori*. The air gap between the vacuum window and the sample is also ill-constrained, affecting absorption at lower energies in particular. For some paint samples there are also sharp changes in the absorption coefficient at elemental absorption edges. Consequently, relative peak intensities contain poorly-defined experimental contributions as a function of energy that may be significant relative to the contribution of the diffracting phases.

Despite these difficulties, it has been found that Rietveld fits of acceptable quality can be achieved by including a scaling factor of the form $a\theta^b$ where $a$ and $b$ are refined parameters. This approach was validated using the NIST $Si_3N_4$ SRM 656 for quantitative phase analysis [42] (see Section S3). For samples that have absorption edges lying within the scanned energy range, the data was divided into segments with each segment having an independent scaling factor of the form described above. Absorption edges are manifested most obviously as sudden jumps in the baseline [25], but in many cases these jumps are very small (e.g. the Ar K-edge at 3206 eV) or not observed at all. In such cases the effect on diffraction peak intensities is negligible and can be ignored. To minimise the number of refined variables, crystallographic parameters affecting intensity (i.e. atomic co-ordinates, site occupancy and atomic displacement parameters) were fixed using published structures. In some cases there were no published displacement parameters in which case generic isotropic values were chosen. Where multiple structures were available, the one with lattice parameters most closely matching the values obtained from the initial Pawley fit was selected. Preferred orientation parameters can also be included in the Rietveld refinements if required.

3. Results

The results of the XRD analyses of the 40 paint diffraction patterns have been categorised into four groups:

   A. Paints for which at least one pigment phase has been identified and, as a minimum, unit cell parameters have been derived.

B. Paints for which only phase(s) other than the pigment has/have been identified (and unit cell parameters derived). Phases such as calcite and barite are commonly used as fillers or extenders.
C. Paints that have clearly identifiable peaks in their diffraction patterns and sufficiently good powder averaging (see below) but incomplete or unsatisfactory analyses. It is possible that these patterns can be analysed if correct phase identifications can be made.
D. The remaining paints have patterns either devoid of any diffraction features at all or have low quality patterns that are not amenable to meaningful analysis.

The number of paints falling into each category are, respectively: 25, four, three and eight, although deciding whether a paint belongs in Group C or D is not always straightforward (see Sections S7.3 and S7.4). Of the eight paints in group D, seven contain synthetic organic pigments. Indeed, no organic pigments have been identified by their diffraction patterns in this work. The results of the analyses are summarised in Table 1.

In total, two of the paints with inorganic pigments produced poor diffraction patterns. The data acquired for the Lamp Black paint is completely devoid of diffraction peaks (Fig. S37h), an unsurprising result given that many carbon-based blacks are amorphous or poorly crystalline [6]. The diffraction pattern of the Viridian paint (pigment: hydrated chromium oxide, $Cr_2O_3 \cdot 2H_2O$) shows very broad peaks (Fig. S36b), consistent with previous studies [43, 44]. Furthermore, the authors could find no published crystal structure for $Cr_2O_3 \cdot 2H_2O$ and it appears that this hydrated oxide is inherently poorly crystalline at best. All of the remaining paints with inorganic pigments have relatively good quality diffraction patterns (groups A and C).

As mentioned in Section 2.2, the degree of powder averaging can be qualitatively assessed by comparing the data from the available detector channels. The majority of the paints show 'moderately good' powder averaging, defined as the observation of the same peaks in the available channels but with varying intensities. A minority of paints have excellent powder averaging with essentially identical peak intensities in all channels (e.g. Fig. S5). The occurrence of peaks appearing in one channel only is observed for perhaps just two or three paints (e.g. Fig. S7); even in these cases there are other phases present with good powder averaging. The main consequence of less than ideal powder averaging is expected to be reduced accuracy in quantitative phase analysis. The observation of reasonably good powder averaging despite the small volume of paint probed by the X-ray beam is perhaps unsurprising given that essentially all paints are prepared by mixing pigments in powdered form with a binding agent.

The crystallographic analyses of several paint samples are highlighted in the sub-sections below, demonstrating the range of information that can be retrieved using the back-reflection EDXRD technique. The data and analyses for all the remaining paints are presented in the Supplementary Information.

### 3.1. Cadmium Gold Yellow and Cadmium Red

The pigments in the Cadmium Gold Yellow and Cadmium Red paints are cadmium sulphide (CdS) and cadmium sulphide selenide ($CdS_{1-x}Se_x$) respectively. CdS can exist in two polymorphic forms: the zinc blende ($F\bar{4}3m$) and the wurtzite ($P6_3mc$) structures, the latter of which is more commonly used in paints thanks to its brighter colour and higher thermal stability [45]. For the solid solution $CdS_{1-x}Se_x$, the full composition range with the wurtzite structure is possible and the shade of red is dependent on the degree of Se substitution [45]. Cadmium yellow is known to degrade in historical paintings, an area of active research [46]. The consensus is that CdS nano-crystallites are more prone to degradation relative to larger crystallites. XRD analysis therefore has a potential role in degradation studies through the identification of crystalline microstructure.

Pawley fits to both diffraction patterns using the wurtzite structure, Fig. 2 (a) and (b), account for all of the observed peaks and the results are reported in Table 2. To determine the Se substitution value for Cadmium Red, literature values for the lattice parameters of wurtzite-type CdS [47], CdSe [48] and intermediate compositions [49-51] were used to calculate the unit cell volume to generate a single metric for comparison. A plot of the unit cell volume as a function of $x$, Fig. 2 (c), shows a near-Vegard's law relationship, yielding $x$ = 0.36 (3) for the phase in the Cadmium Red paint sample. It is assumed that no other elements are involved in the solid solution.

The diffraction peak widths and profile shapes show some interesting behaviour. The (103) and (203) reflections, and only these two, have 'super-Lorentzian' lineshapes, while the adjacent (210) and (211) peaks have asymmetric profiles with significant intensity appearing between these peaks. These distinct features are present in both diffraction patterns but to a lesser extent in the Cadmium Red pattern. In addition, the (110) peak is moderately asymmetric in the Cadmium Gold Yellow pattern only. Previous studies of CdS nanoparticles have commonly found that the (103) peak is especially broad [52-55]. The anisotropy in the peak widths as a function of $hkl$ as exhibited by the Cd pigments is characteristic of stacking faults which are commonly found in hexagonal crystal structures [56]. Warren [57] has shown that reflections with $h - k = 3n$ (where $n$ is an integer) are unaffected and remain sharp whereas other reflections are broadened as a consequence of stacking faults, proportional to both the fault density and the magnitude of $l$. Quantitative analysis of the microstructure of these pigments is beyond the scope of this study.

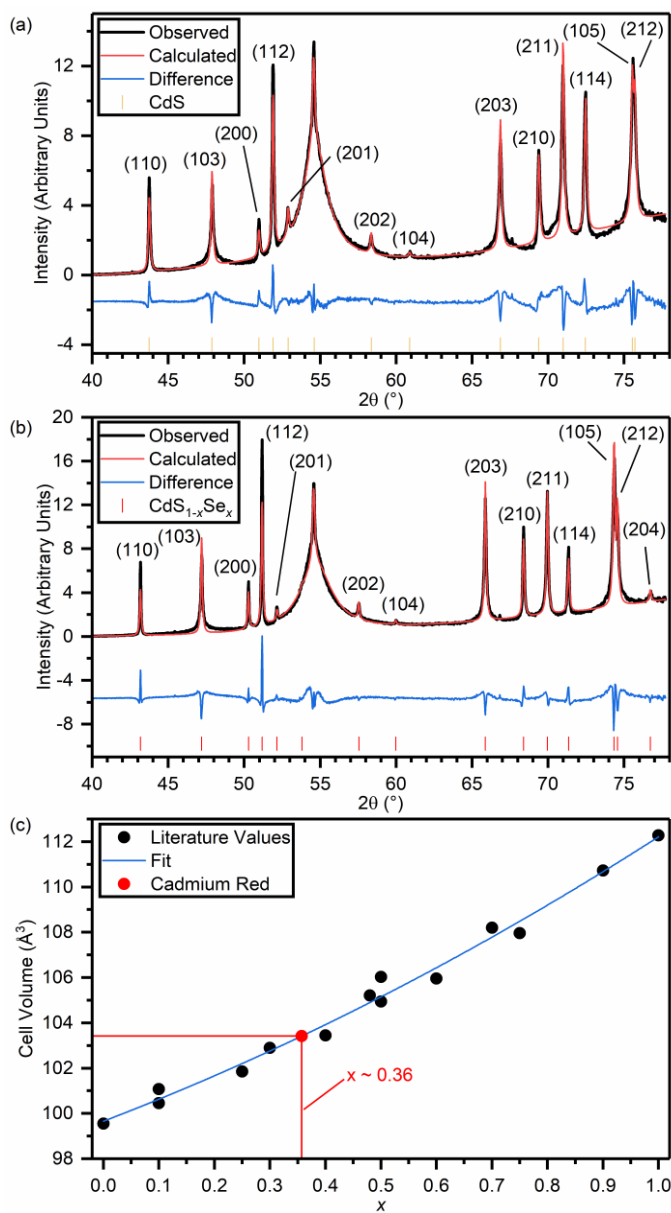

Fig. 2. Pawley fits to the (a) Cadmium Gold Yellow and (b) Cadmium Red diffraction patterns. The broad peak centred on $2\theta$ = 54.8° is due to diffraction by the graphitic vacuum window, and was fitted by inclusion of two pseudo-Voigt functions. The difference plots are shown offset on the vertical scales. (c) Plot of the unit cell volume of $CdS_{1-x}Se_x$ solid solutions vs $x$ from literature data with a quadratic fit, together with an estimate of $x$ for Cadmium Red.

**Table 2.** Crystallographic parameters of the identified phases in the Cadmium Gold Yellow and Cadmium Red paints.

| Phase | Sowa [47] | Pawley Fit |
|---|---|---|
| **CdS ($P6_3mc$)** | | |
| $a$ (Å) | 4.1365 (3) | 4.13435 (7) |
| $c$ (Å) | 6.7160 (4) | 6.7159 (2) |
| $R_{wp}$ % | | 11.6 |
| **CdS$_{1-x}$Se$_x$ ($P6_3mc$)** | | |
| $a$ (Å) | | 4.18758 (6) |
| $c$ (Å) | | 6.81004 (14) |
| $x$[a] | | 0.36 (3) |
| $R_{wp}$ % | | 17.0 |

[a]This parameter was not derived in the Pawley fit; see main text for details.

### 3.2. Manganese Violet

The pigment in Manganese Violet is a manganese complex, $NH_4MnP_2O_7$, and two polymorphs have previously been demonstrated to be present in commercial paint samples [58, 59]. Past work with non-invasive portable XRD has struggled to assign peaks in Manganese Violet samples due to the complex diffraction pattern and the limited resolution and $d$-space range afforded by the instrument [59]. The resolution of the back-reflection EDXRD experimental technique yields a high-quality diffraction pattern from the Manganese Violet paint, to which a two-phase Pawley fit gives excellent agreement. It has been assumed that the broadening of peaks due to the sample, beyond instrument broadening, is caused by small crystallite domain sizes. On this basis, the volume-weighted mean crystallite diameters of the α- and β-polymorphs are 102 (3) nm and 480 (40) nm respectively. The relatively large crystallite size, and the large error, derived for the β-polymorph should be interpreted in the context of the limit of applicability of the Scherrer equation that is generally on the order of 100 – 200 nm [60, 61]. The refined lattice parameters of both polymorphs (Table 3) are close to published values [58], supporting the use of the published crystal structures in the subsequent Rietveld fit. The phase weight fractions are the same within errors to the values obtained by Begum and Wright [58] even though the paint they analysed came from a different supplier, possibly suggesting that both paint suppliers procured the pigment from the same manufacturer. It is notable that these authors were not able to synthesise the β-polymorph in isolation.

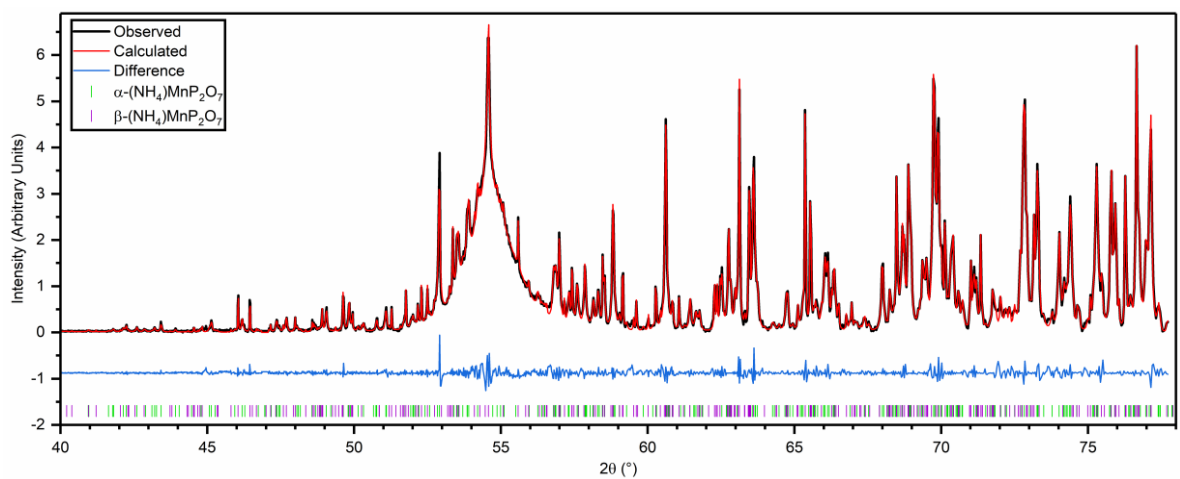

**Fig. 3.** Pawley fit to the Manganese Violet diffraction pattern. The broad peak centred on $2\theta = 54.8°$ is due to diffraction by the graphitic vacuum window, and was fitted by inclusion of two pseudo-Voigt functions. The difference plot is shown offset on the vertical scale.

**Table 3.** Crystallographic parameters of the identified phases in the Manganese Violet paint.

| Phase | Begum and Wright [58] | This work |
|---|---|---|
| **α-NH$_4$MnP$_2$O$_7$ ($P2_1/c$)** | | |
| $a$ (Å) | 7.4252 (3) | 7.4150 (2) |
| $b$ (Å) | 9.6990 (4) | 9.7047 (3) |
| $c$ (Å) | 8.6552 (4) | 8.6451 (3) |
| $β$ (°) | 105.627 (3) | 105.551 (2) |
| Weight fraction[a] (%) | 69 (3) | 71 (2) |
| Mean crystallite size (nm) | | 102 (3) |
| **β-NH$_4$MnP$_2$O$_7$ ($P\bar{1}$)** | | |
| $a$ (Å) | 8.4034 (6) | 8.40359 (14) |
| $b$ (Å) | 6.1498 (4) | 6.14997 (5) |
| $c$ (Å) | 6.1071 (4) | 6.10641 (6) |
| $α$ (°) | 104.618 (5) | 104.6288 (11) |
| $β$ (°) | 100.748 (5) | 100.7411 (10) |
| $γ$ (°) | 96.802 (6) | 96.8044 (11) |
| Weight fraction[a] (%) | 31 (3) | 29 (2) |
| Mean crystallite size (nm) | | 480 (40) |
| $R_{wp}$ (%) (Pawley/Rietveld) | | 11.3/33.1 |

[a]Derived from the Rietveld refinement; all other parameters are from the Pawley fit.

### 3.3. Flake White and Cremnitz White

Lead white is a pigment widely used since antiquity and well into the 20$^{th}$ century despite the advent of cheaper and safer whites, the oxides of zinc and titanium, developed in the 19$^{th}$ and 20$^{th}$ centuries respectively [6]. Lead white can contain a range of lead carbonate compounds [6], usually dominated by lead carbonate hydroxide [2PbCO$_3$•Pb(OH)$_2$, hydrocerussite] and lead carbonate (PbCO$_3$, cerussite). The ratio of these two phases, and their crystallite sizes, shapes and preferred orientation, can affect the optical properties of the paint, as well as being indicative of synthesis method and painters' techniques [5, 62-67]. Powder XRD analysis offers the most powerful method to distinguish the lead carbonates and elucidate their structural properties in paints.

The Flake White and Cremnitz White paints analysed in this study are both based on lead white. Both contain cerussite and hydrocerussite while Flake White also contains ZnO. The unit cell parameters of each phase were derived from Pawley fits to the data and are close to literature values (Table 4). A series of Pawley fits were also used to investigate the microstructure of the phases. Hydrocerussite commonly occurs as platy crystallites [65] and the observed anisotropic peak broadening was fitted using the crystallite shape model of Ectors *et al* [40] assuming a cylindrical shape. The results suggest that the hydrocerussite crystallites in Cremnitz White are significantly more anisotropic in shape than in Flake White (Table 4). It should be noted that trial Pawley fits using the Stephens anisotropic strain model [68] gave comparable $R_{wp}$ values, and the presence of microstrain in the hydrocerussite phases (with or without size effects) cannot be ruled out based on these results.

Attempts to fit the anisotropic shape model to the cerussite phases gave unconvincing results and isotropic shapes were assumed in the final Pawley fits. In both paints the cerussite phase exhibits poor powder averaging, relative to the hydrocerussite and ZnO phases, and in some instances the peak shapes are irregular, suggesting the influence of larger, individual crystallites. For this reason, the crystallite sizes derived for the cerussite phases are not considered reliable and this parameter has only been included in the refinements to improve the overall fit. An anisotropic shape model was also not supported for the ZnO phase in Flake White. Rietveld refinements were performed using published crystal structures (as reported in Table 4) in order to derive phase ratio quantifications. The hydrocerussite phase in both paints shows preferred orientation in the <001> direction, as expected for the platy crystallites. The degree of preferred orientation is significantly greater for Cremnitz White, consistent with the more pronounced shape anisotropy. For Flake White, introducing the single preferred orientation parameter, $r$, of the March-Dollase model [69] gave an adequate fit with most of the discrepancy between the fit and the data arising from the poor powder averaging of the cerussite phase, Fig. 4(b). A corresponding fit to Cremnitz White gives a much lower value of $r$ (Table 4), indicative of greater preferred orientation, but the $R_{wp}$ factor is relatively high and there remain significant errors in the fit to the hydrocerussite peak intensities. Instead, an eight-term spherical harmonic [70] was used to model the hydrocerussite preferred orientation. The remaining discrepancies are again primarily due to cerussite poor powder averaging. The latter factor undoubtedly limits the accuracy of the quantification (see Section S6). In both paints, hydrocerussite is the dominant lead carbonate phase as commonly observed in Old Masters paintings [65], for example, and consistent with the stack manufacturing process [66].

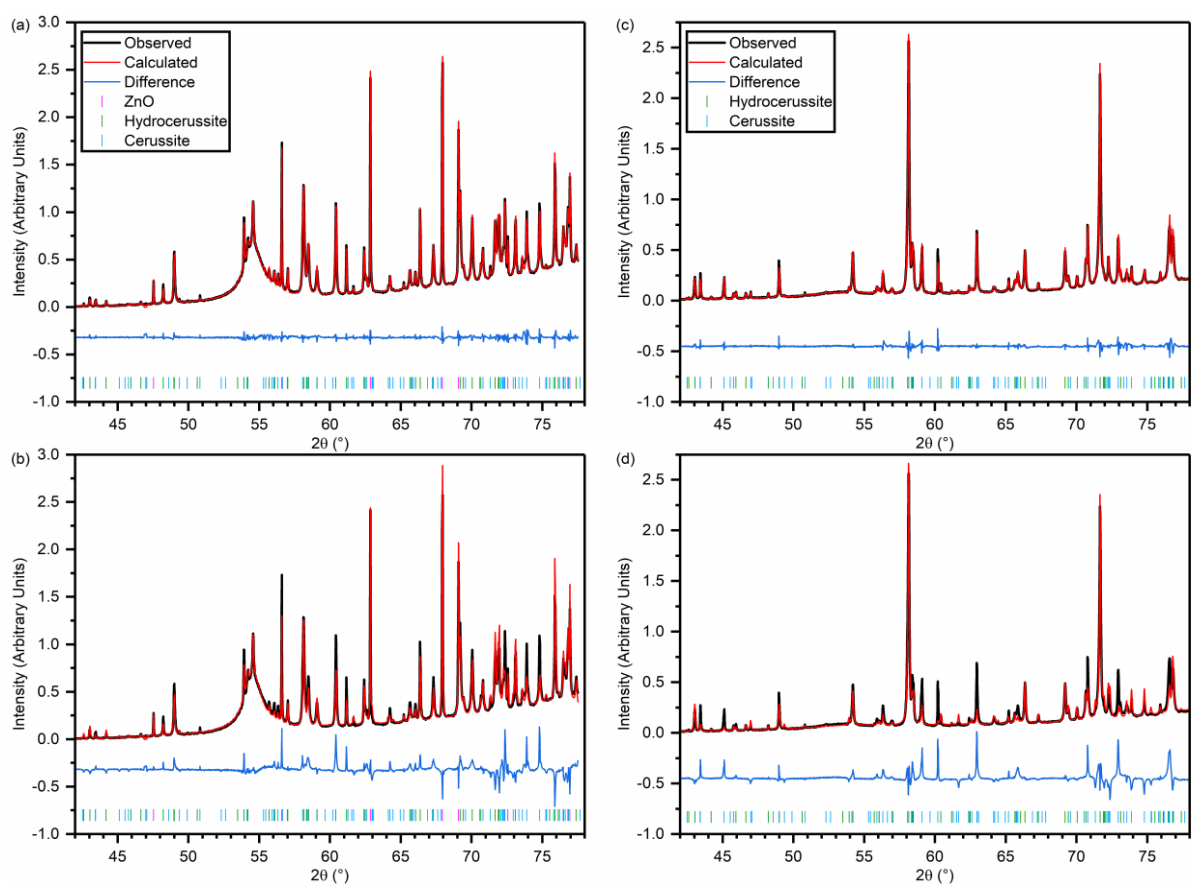

**Fig. 4.** (a) Pawley fit and (b) Rietveld refinement to the Flake White diffraction pattern. (c) Pawley fit and (d) Rietveld refinement to the Cremnitz White diffraction pattern. In each case the difference plot is shown offset on the vertical scale.

Table 4. Crystallographic parameters of the identified phases in the Flake White and Cremnitz White paints.

| Phase | Literature | Flake White | Cremnitz White |
|---|---|---|---|
| **$2PbCO_3 \cdot Pb(OH)_2$ ($R\bar{3}m$)** | Siidra et al [71] | | |
| $a$ (Å) | 5.2475 (1) | 5.24405 (2) | 5.24372 (6) |
| $c$ (Å) | 23.6795 (7) | 23.6819 (2) | 23.68405 (10) |
| $L_x$ (nm) | | 136 (6) | 230 (30) |
| $L_z$ (nm) | | 32.6 (6) | 32.8 (4) |
| $L_x/L_z$ | | 4.2 (2) | 7.0 (8) |
| Weight fraction[a] (%) | | 56.2 (10) | 81.0 (8) |
| $r$ (001)[a] | | 0.709 (6) | 0.383[b] (3) |
| **$PbCO_3$ (*Pmcn*)** | Antao & Hassan [72] | | |
| $a$ (Å) | 5.18324 (2) | 5.18236 (13) | 5.18265 (5) |
| $b$ (Å) | 8.49920 (3) | 8.4968 (2) | 8.50094 (9) |
| $c$ (Å) | 6.14746 (3) | 6.1435 (2) | 6.14231 (8) |
| Crystallite size (nm) | | 95 (2) | 89 (2) |
| Weight fraction[a] (%) | | 10.9 (5) | 19.0 (8) |
| HC/(HC + C)[c] (%) | | 83.8 (11) | 81.0 (8) |
| **ZnO ($P6_3mc$)** | NIST [38] | | |
| $a$ (Å) | 3.24983 (8) | 3.249919 (9) | |
| $c$ (Å) | 5.2068 (1) | 5.20682 (3) | |
| Weight fraction[a] (%) | | 32.9 (9) | |
| $R_{wp}$ (%) (Pawley/Rietveld) | | 4.7/12.1 | 7.2/21.0 |

[a]Derived from the Rietveld refinements; all other parameters are from the Pawley fits.
[b]This value is derived from a separate Rietveld refinement in order to have a value directly comparable with the Flake White result.
[c]HC = hydrocerussite, C = cerussite.

### 3.4. Naples Yellow

Naples Yellow is commonly identified as lead antimonate ($Pb_2Sb_2O_7$) which has a cubic pyrochlore structure. There are several closely-related yellow lead-based mixed oxide pigments which have a complicated history in glass-making, ceramics and paintings, having been lost and rediscovered several times [73-75]. Remarkably, the occurrence of the lead tin antimonate phase $Pb_2SbSnO_{6.5}$ in pre-20[th] century paintings was rediscovered as recently as 1998 [76, 77]. Furthermore, doubt has been cast on the existence of stoichiometric $Pb_2Sb_2O_7$ as a cubic phase in a recent study suggesting that partial substitution of Pb with Na is required for stabilisation [74]. Further art historical research into the relationship between the characteristics of these pigments and their temporal and geographical use offers considerable potential for authentication and provenancing purposes [73, 74].

The Naples Yellow pattern, Fig. 5, has several broad diffraction features (referred to here as 'bands') with relatively steep sides, behaviour that can be recognised as characteristic of a continuous series of phases within a solid solution [78]. The positions of these bands are consistent with cubic pyrochlore phases. A Pawley fit of the pattern shows that the end-members have lattice parameters of $a = 10.39413$ (6) Å and 10.56866 (8) Å, comparable to literature values for $Pb_2Sb_2O_7$ (10.40 Å [79]) and $Pb_2SbSnO_{6.5}$ (10.5645 Å [80]). To adequately fit the intensity between the end-member phase peaks, a total of 12 additional pyrochlore phases were included with their lattice parameters constrained to spread evenly between the end-member values. This number of phases is required to model the steep sides of the diffraction bands and to capture the relatively sharp peaks within some of the bands. Barite ($BaSO_4$, *Pnma*) has also been identified and is included in the Pawley fit [unit cell parameters: $a = 7.15335$ (6) Å, $b = 8.87463$ (8) Å, $c = 5.45266$ (6) Å]. It is apparent from inspection of the data from the four SDD channels that the powder averaging of this sample is relatively poor (Fig. S8) and a Rietveld refinement to quantify the phase ratios has not therefore been attempted.

Several previous XRD and Raman studies of lead antimonate pigments have observed bimodal compositions, both in artworks and in synthesised samples [73, 74, 81-83], though with lattice parameters that do not necessarily correspond to the stoichiometric end-members. In contrast to the data presented here, these studies appear to show a genuine bimodality rather than a continuous solid solution. This disparity is presumed to relate to differences in the pigment production methods.

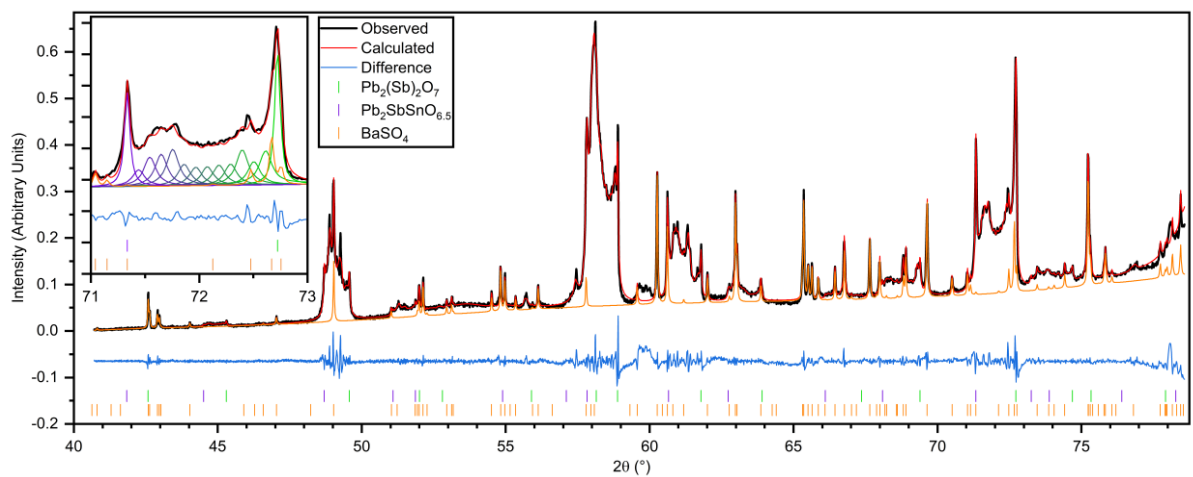

**Fig. 5.** Pawley fit to the Naples Yellow diffraction pattern. The fitted barite peaks are shown in orange. The inset shows an expansion of the $2\theta = 71° – 73°$ range with all of the fitted peaks due to the solid solution phases shown colour-coded. The difference plot is shown offset on the vertical scale.

## 4. Discussion

A significant outcome of this study is the lack of detection of any of the organic pigments, even those containing a heavy element such the copper phthalocyanine pigments. Some weak peaks are visible in some of the diffraction patterns of paints containing an organic pigment (see Fig. S37), but nothing that allows identification. Based on this work alone, it is not possible to determine whether the failure to detect organic pigments results from a lack of sensitivity or because they are poorly crystalline or amorphous.

Organic pigments typically have high tinting strength [84] and therefore constitute a small weight fraction of commercial paints. Organics are weak X-ray scatterers because of their low average atomic number and paints based on organic pigments frequently contain inorganic fillers and extenders [85]. (In principle, if the XRD characteristics of the inorganic components associated with a specific organic pigment are sufficiently distinct, and the association can be made unambiguously, the inorganics could act as tracer for the organic pigment. It seems likely that these criteria will rarely be met in practice however). The inorganic phases are much more efficient diffractors and produce patterns that may overlay the organic diffraction peaks. More importantly, they limit the penetration depth of X-rays into the paint by absorption and therefore significantly reduce the probed volume and hence the diffraction signal from the organic components. Furthermore, previous studies have shown that dry organic pigment powders typically produce clearly identifiable diffraction patterns only below $2\theta \approx 30°$ (Cu-K$\alpha$) [9, 84, 86], equivalent to $d = 3.0$ Å or 2.1 keV in back-reflection EDXRD. As currently implemented, the back-reflection technique cannot access this range. All these factors, acting in concert, predict weak diffraction from organic pigments, perhaps below the limit of sensitivity in the present study. Additionally, the extent to which organic pigments retain crystallinity when prepared as paints is unclear. It is interesting to note that two earlier studies concluded that organic pigments are more difficult to identify with XRD in oil-based paints compared to acrylic or alkyd binders, though it is not clear whether the primary factor is the more frequent presence of inorganic components [85] or poor crystallinity [9]. In future studies it will be interesting to see whether diffraction patterns can be observed for organic pigments in acrylic or alkyd media and whether larger $d$-spacing peaks can be detected when extending the method to lower energies (see below).

The current limitation on the observable $d$-spacing range can affect the analysis of inorganic paints also. Large $d$-spacing/low-angle diffraction peaks are highly diagnostic for phase identification [25] and the absence of these peaks in the data reported here can hinder identification. For most of the paints, the identity of the pigment and any extenders was either known in advance or could be restricted to a limited set of candidates. One exception is the Terre Vert paint (Fig. S36a) which contains the green earth pigment, described by the supplier as a complex of ferrous silicates in aluminium and magnesium clays. Green earth most commonly contains the natural green clay minerals glauconite and celadonite [6] but there are exceptions [87] and modern varieties may be adulterated with synthetic organic greens [88]. To date, the authors have been unable to find satisfactory identifications of the phases present in the Terre Vert paint.

The results of this study demonstrate the range of information about the paint samples that can be extracted using the back-reflection EDXRD technique, non-invasively and at high resolution: phase identification (including solid solutions), highly accurate unit cell parameters, quantification of the observed phases, crystallite size and strain parameters and preferred orientation parameters. The paints chosen for presentation in Section 3 highlight these capabilities. The high resolution of the method enables effective analysis even in cases of very high peak density such as the Manganese Violet paint which contains a pigment with two low-symmetry polymorphs. In all of the paint data acquired in this study, the widths of the diffraction peaks are limited by sample effects, not by instrumental broadening. Although Manganese Violet is not a particularly common pigment, the other pigments for which results are presented in Section 3 are all highly significant in art history and conservation and are the subjects of active research. For example, there is considerable interest in the microstructural properties of the lead white phases and how they relate to historical production methods [62, 64-67]. The analysis of microstructure by powder XRD is a very broad topic within the field, ranging from basic application of the Scherrer equation to modelling of the whole pattern using highly detailed physical models such as the work of Scardi and Leoni [89-91], and there is a concomitant requirement of time and effort to achieve meaningful results. In complex cases, depending on the aims of the study, the effort required for a detailed physically-based analysis may not be warranted. However, in such cases the observed pattern of diffraction peak widths and profile shapes can serve as a fingerprint that may be unique to a specific artist or supplier within a limited geographical region and timeframe. The cadmium pigments reported in Section 3.1 provide a good example. Once again, the high resolution of the back-reflection EDXRD technique is crucial in faithfully recording the characteristics of the diffraction pattern determined by the sample physical properties and not limited by instrumental effects.

In general terms, phase identification can contribute to art historical knowledge for artworks of known provenance, or aid attribution and authentication where provenance is in doubt. The distinction and quantification of polymorphs (e.g. Manganese Violet) and other closely-related phases (e.g. lead tin antimonates in Naples Yellow) allows the synthesis of more in-depth knowledge than pigment identification alone. As mentioned in connection with lead white, phase quantification along with crystallite size and shape information can distinguish alternative production methods [65] which, again, can either yield historical insights or serve to delimit the geotemporal production. There is also considerable potential to discern details of the artist's technique, whether through the choice of specific pigment variants [5] or the application of paint onto a canvas deduced by mapping preferred orientation

[92]. Although not demonstrated as part of this study, there is every reason to suppose that degradation phases can be detected using the back-reflection method, allowing elucidation of degradation mechanisms and informing the development of conservation and restoration strategies.

This study has focused on the analysis of individual paint samples and, clearly, real paintings will present additional challenges. The dilution and mixing of paints will reduce the diffraction signal from any one phase and yield more complex diffraction patterns. The high resolution of the technique is a major advantage in mitigating the challenge posed by complexity. The presence of amorphous varnish overlaying the paint will also attenuate the diffraction signal of the pigments. A possible way to mitigate this factor is to extend the method to higher X-ray energies to take advantage of their greater penetrating power (see Section S5). EDXRD patterns become more crowded at higher energies [23] but, again, high resolution is beneficial in this respect. It is possible that, with the appropriate calibrations, the thickness of a varnish layer could be estimated via the attenuation effect as a function of energy. In any case, the demonstrated performance of the back-reflection EDXRD technique presented here establishes a baseline of what is achievable in favourable cases. Additionally, it is possible in principle to retrieve the relative depths of the diffracting phases within the probed volume by assessing diffraction peak intensity ratios as a function of X-ray energy, as long as the phases show good powder averaging. In this way, it may be possible to distinguish paint layers.

The need to access synchrotron facilities is a disadvantage in cultural heritage studies because of the limited beamtime available and the need to transport valuable objects. It is possible to transfer the back-reflection technique into the laboratory with only a modest penalty in terms of resolution through the use of transition-edge sensor (TES) arrays [93-95], also known as microcalorimeters, a technology that has undergone rapid development in recent years. These devices are state-of-the-art X-ray detectors offering spectral resolutions on the order of a few eV, approaching two orders of magnitude improvement over the best solid state detectors [96]. The development of sensor arrays and advanced sensor design and signal processing methods allows detection rates in the 10s kilo-counts per second range, with greater capabilities expected in the future [94]. These rates are compatible with the requirements of back-reflection EDXRD using laboratory sources [23].

The use of microcalorimeters in a laboratory setting brings multiple advantages. The most significant advantage, relative to synchrotron implementation, is the simultaneous acquisition of the whole diffraction pattern. It is this factor that allows the use of a laboratory source despite the orders of magnitude reduction in brilliance, also reducing the risk of radiation damage. Implementation with microcalorimeters will also allow access to lower X-ray energies/larger $d$-spacings, subject to the limitations imposed by window and air absorption if samples are mounted in free-space. Where necessary, objects can be mounted within a bagged enclosure flushed with He. For objects or artworks that exhibit poor powder averaging, the deleterious effect on diffraction patterns can be mitigated by employing a larger beam spot; for example, a spot size of 10 mm will increase the probed volume by a factor of 100 relative to the ~1 mm spot size employed in the synchrotron experiments (assuming sample homogeneity). Further improvement is achievable via lateral scanning and/or sample tilting during data acquisition. For broadband detection the sample movement can be *slow*, whereas for the equivalent averaging effect at the synchrotron the same movement would have to be repeated at each step during scanning of the X-ray energy. One final advantage of microcalorimeters in this context will be the simultaneous acquisition of highly complementary XRF data, yielding elemental composition. The occurrence of XRF peaks in EDXRD is normally considered a disadvantage because of the potential for overlap [97], but with eV-scale resolution this factor becomes an advantage. It may even be possible in some instances to deduce chemical environment data from XRF data [95, 98], in support of phase identification through XRD analysis. The authors anticipate that a microcalorimeter-based instrument will be transportable [Joseph Fowler, NIST, personal communication, 2019], allowing *in situ* investigations in museums and art galleries and access to artefacts that cannot be moved because of their bulk or prohibitive insurance costs.

On the basis of acquisition time per measurement spot of ~20 mins, 2D mapping [99] of a painting would take a prohibitively long time to record a meaningful dataset. If the primary aim was simply to identify the pigments present in the mapped area, the scan time per spot could be reduced significantly by scanning the monochromator more quickly and over a shorter energy range. It would be essential to ensure that the more intense, resolved diffraction peaks of pigments of interest are within the scanned range. Even with these changes, it is likely that measurement times would still need to be a few mins per spot whereas dwell time in the secs range are required for mapping applications [99]. When implemented using microcalorimeters, simulations suggest that acquisition times can be *shortened* relative to the synchrotron, though the improvement factor depends strongly on details of the experimental design such as the geometry and X-ray tube power. Insufficient work has been performed to date to assess potential acquisition times and whether mapping applications will be feasible. Longer total acquisition times can be tolerated for a laboratory instrument, favouring the possibility of mapping applications. Good spatial resolution can be achieved with the use of a focusing polycapillary optic [100] to give a small beam spot on the sample.

As with all analytical techniques, the back-reflection EDXRD method comes with a set of advantages and disadvantages. The key advantage, indeed the *raison d'etre* of the method, is that it is extremely tolerant of the surface morphology of the sample [23,25]. It follows that there is also no need to position samples a precise distance relative to the instrument. These advantages hold true even when implemented at the very high resolution afforded at the synchrotron; samples can be moved several mm away from the instrument without significant consequence [25], an inconceivably large distance by the standards of conventional angle-dispersive XRD. Furthermore, we fully expect to be able to transfer the technique into the laboratory and museum using microcalorimeter technology with only a modest penalty in the achievable spectral resolution. The results of a direct comparison between synchrotron and laboratory data for two of the paints (Cobalt Blue and Cremnitz White) are shown in Section S6 to illustrate the superior resolution of the EDXRD data. High resolution enables the extraction of an extensive set of crystallographic parameters, including microstructure, contrasting with current portable XRD instrumentation (see sec. 1). A key part of our plans includes proof-of-principle experiments using microcalorimeters. An additional advantage of the method is the convenience of the inherent back-reflection geometry that minimises the potential for physical interference between the instrument and the object being studied.

One of the disadvantages of the technique is the difficulty in accessing low energies/large *d*-spacings. On beamline B18 at Diamond, there is a hard limit of 2.05 keV set by the Si(111) monochromator, but more generally absorption in air and by the vacuum window becomes a limiting factor at these soft X-ray energies. Quantitative phase analysis of EDXRD data is more challenging than for angle-dispersive methods because of the ill-constrained energy-dependence of absorption and other parameters, leading in general to reduced accuracy. Also, achieving high resolution requires either transportation of artefacts to a synchrotron facility or the use of expensive microcalorimeters. It is anticipated that an instrument based on microcalorimeters will be transportable, but certainly considerably bulkier than existing portable XRD equipment. Good powder averaging cannot be guaranteed for any specific sample, but this issue applies to all non-invasive XRD techniques.

In summary, we believe that the combination of non-invasive capability and uncompromised data quality through high spectral resolution distinguishes the back-reflection EDXRD technique from alternative non-invasive XRD methods.

## 5. Conclusions

The back-reflection EDXRD technique, implemented at high resolution, shows excellent potential for the analysis of inorganic pigments in paints. Of the 40 paints for which data was acquired, 29 are inorganic-based and successful analysis was possible for 25 of these (86%), where success is defined as phase identification and retrieval of the unit cell parameters of the pigment. With additional effort, two or three further paints may become analysable. In contrast, none of the organic pigments were detected though it is worth checking in the future whether acrylic- or alkyd-based organic pigments are detectable, especially when access to larger $d$-spacings is possible. It is noted that non-invasive Raman spectroscopy is effective for organic pigments [9], for identification purposes at least, and in this respect there is good complementarity between XRD and Raman methods. Synthetic organic pigments were developed from the first half of the 20$^{th}$ century onwards and, consequently, it is expected that the back-reflection method is most suited to the archaeometric study of older paintings.

The efficacy of the technique for real paintings, with the complicating factors of varnish overlayers, paint mixtures and layering of paint, requires testing. The utility of current non-invasive XRD methods within cultural heritage, along with the quality of results presented in this article, provides a foundation for optimism. Systematic studies with prepared materials are planned in order to test the potential for depth analysis and to quantify the effect of varnish.

It is emphasised that high resolution back-reflection EDXRD is capable of returning very high quality crystallographic analyses entirely non-invasively. No special attention needs to be paid to the separation between the instrument and the sample other than to keep the air gap as small as possible and there is no need to consider the surface morphology of the sample. Non-invasive analyses preserve the integrity of heritage artefacts and allows more representative assessment of an object by enabling more thorough measurement sampling. Adapting analytical techniques for non-invasive application generally involves a technical compromise of some description, leading to reduced data quality. The exciting potential of the back-reflection EDXRD method within the cultural heritage field in general, and paintings in particular, lies in this lack of compromise. High quality non-invasive analysis is the holy grail of archaeometry.

### Acknowledgements

We acknowledge Diamond Light Source for time on B18 under proposals SP18533 and SP23341, and in particular Dr Giannantonio Cibin for extensive support during beamtime. We thank the Engineering and Physical Sciences Research Council for funding (standard grant number EP/R024626/1).

# High-Resolution Non-Invasive X-ray Diffraction Analysis of Artists' Paints: Supplementary Material

1. **Test Panel**

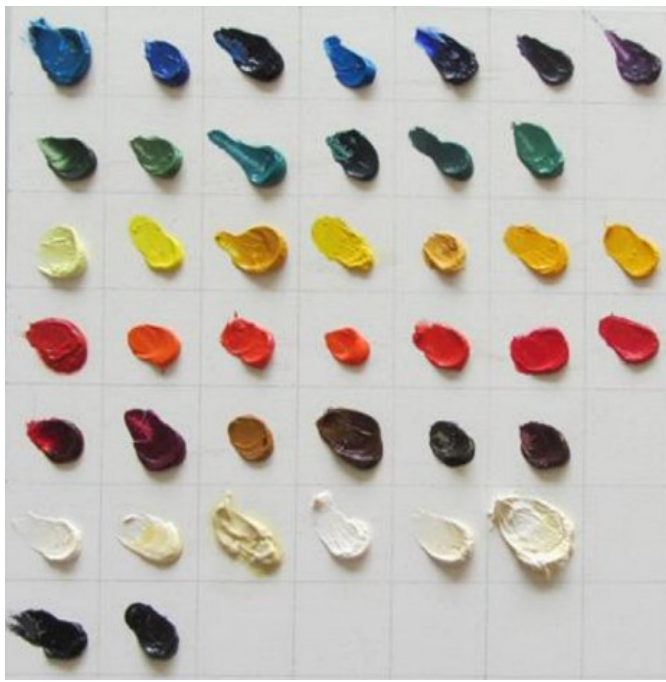

**Fig. S1.** Photograph of the panel of test paints. An EDXRD spectrum of the paint on the canvas was also acquired – this paint contains the titanium white pigment (TiO$_2$, rutile).

**Table S1.** Grid showing the identification of each paint (supplier's name), matching the layout in Fig. S1.

| Cerulean Blue | Cobalt Blue | Phthalocyanine Blue Lake | Manganese Blue | Ultramarine Blue | Dioxazine Violet | Manganese Violet |
|---|---|---|---|---|---|---|
| Terre Vert | Chrome Oxide Green | Cobalt Turquoise | Phthalocyanine Green Lake | Viridian | Chrome Green | |
| Lemon Yellow | Bright Yellow Lake | Aureolin | Yellow Lake | Naples Yellow | Cadmium Gold Yellow | Chrome Yellow |
| Vermilion | Minium | Orange Molybdate | Pyrrolo Vermilion | Naphthol Red | Scarlet Lake | Cadmium Red |
| Alizarin Crimson | Magenta | French Yellow Ochre | Raw Sienna | Raw Umber | Transparent Oxide Red | |
| Cremnitz White | Flake White | Barite White | Zinc White | Flemish White | Titanium White No. 3 | |
| Ivory Black | Lamp Black | | | | | |

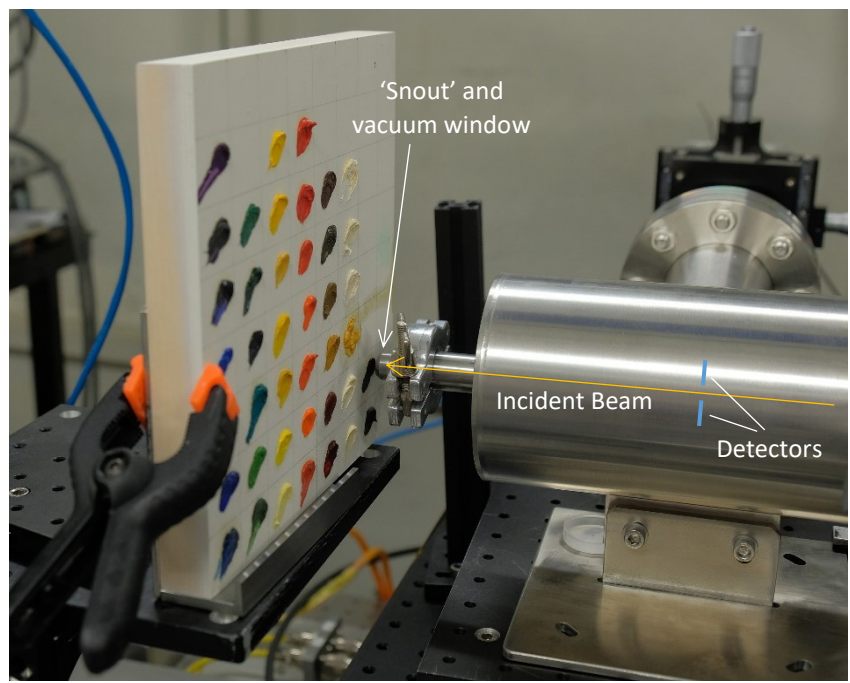

**Fig. S2.** Photograph of the test panel mounted in position for a measurement. The panel was secured on an *xyz*-translation stage and positioned so that each paint was within 2 – 3 mm of the vacuum window at the end of the 'snout'. The location of the detectors within the vacuum chamber is indicated on the photograph (not to scale).

## 2. Calibration of the February 2019 Data

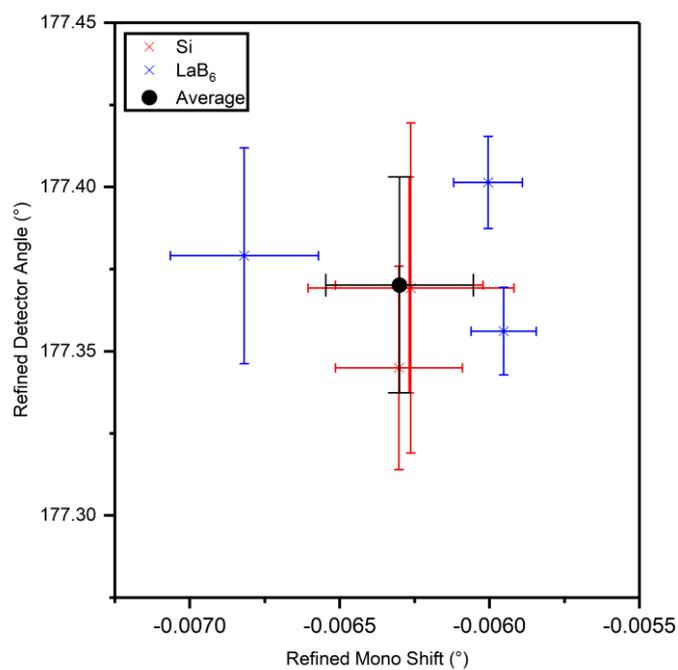

**Fig. S3.** Calibration of $\delta\theta$ and $\delta\theta_{DCM}$ using the LaB$_6$ and Si powder SRMs for the February 2019 beamtime data (see Section 2.3.1 of the main article).

## 3. Validation of Phase Quantification

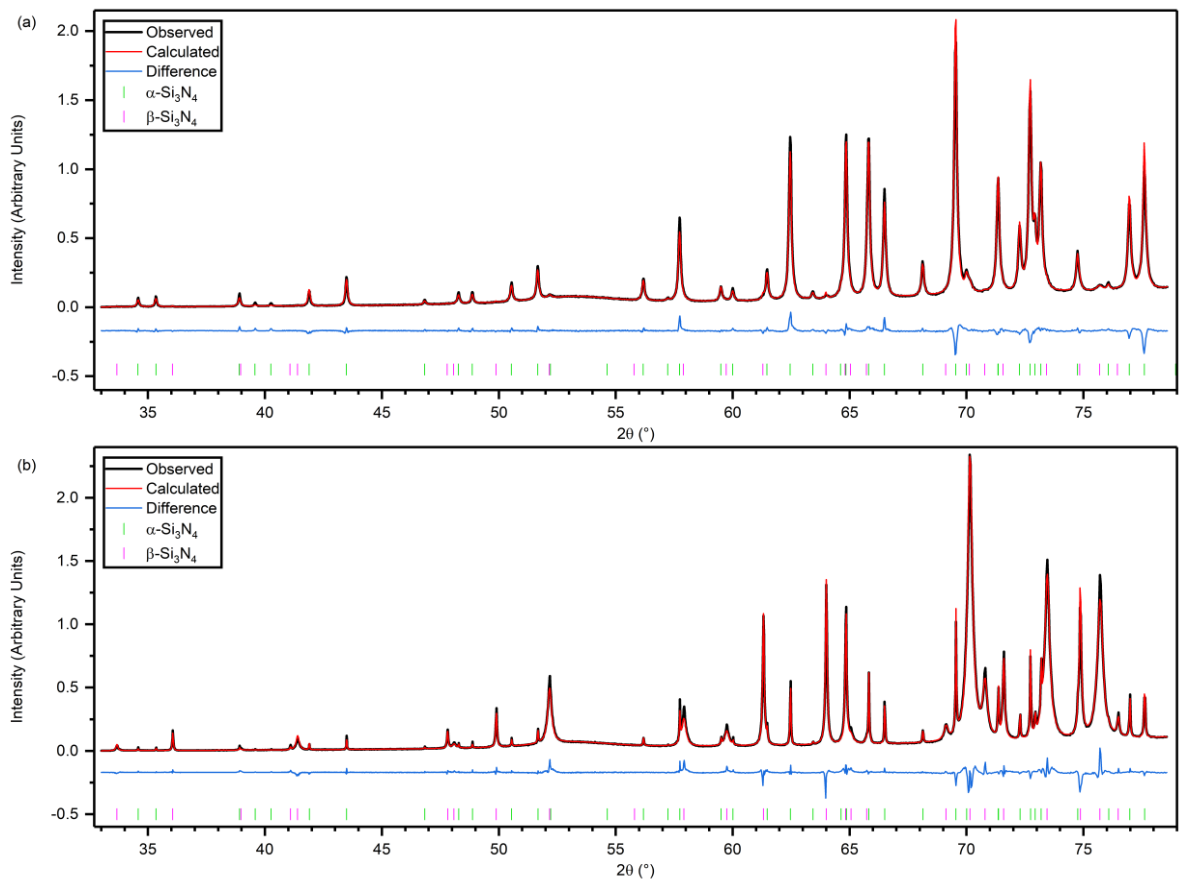

**Fig. S4.** Rietveld refinements of the NIST 656 SRM data: (a) α 656, (b) β 656.

**Table S2.** Comparison of the NIST certified quantification of α- and β-$Si_3N_4$ in SRM 656 with fitted values in this work.

| Standard | NIST Certification* | Rietveld Fit |
|---|---|---|
| **α 656** | | |
| α (%) | 96.7 (1) | 97.0 (2) |
| β (%) | 3.3 (1) | 3.0 (2) |
| $b^†$ | | 3.37 (3) |
| $R_{wp}$ (%) | | 6.55 |
| **β 656** | | |
| α (%) | 17.8 (3) | 13.3 (7) |
| β (%) | 82.2 (3) | 86.7 (7) |
| $b^†$ | | 3.59 (2) |
| $R_{wp}$ (%) | | 7.50 |

*The NIST quantification includes amorphous content. The crystalline component quantities have been normalised to 100% in this table.
†Parameter from the $a\theta^b$ scaling factor. The $a$ parameter is the overall linear scaling parameter with arbitrary units.

The Rietveld refinements of the NIST 656 SRM data acquired during the January 2020 beamtime are shown in Fig. S4 and the quantitative results are reported in Table S2. The values for the α 656 component agree within errors while there is a discrepancy of 4.5% in the quantification of β 656. The powder averaging for both components is good but is somewhat worse for β 656 and this factor provides a possible explanation for the error in quantification. It is also possible that the $a\theta^b$ scaling factor is failing to accurately account for the intensity variation as a function of X-ray energy. The magnitude of the discrepancy provides a rough estimate of the likely uncertainties in the phase quantifications of the paint samples.

## 4. Assessment of Powder Averaging

As described in Sections 2.2 and 3 of the main article, the degree of powder averaging can be assessed qualitatively by comparison of the diffraction peak intensities in each of the detector channels (two working detectors in the February 2019 beamtime and four in January 2020). The figures below (S4 – S7) illustrate the behaviour ranging from good powder averaging to poor. Naturally, the assessment is better with four available detectors and all the examples have been chosen from the January 2020 beamtime.

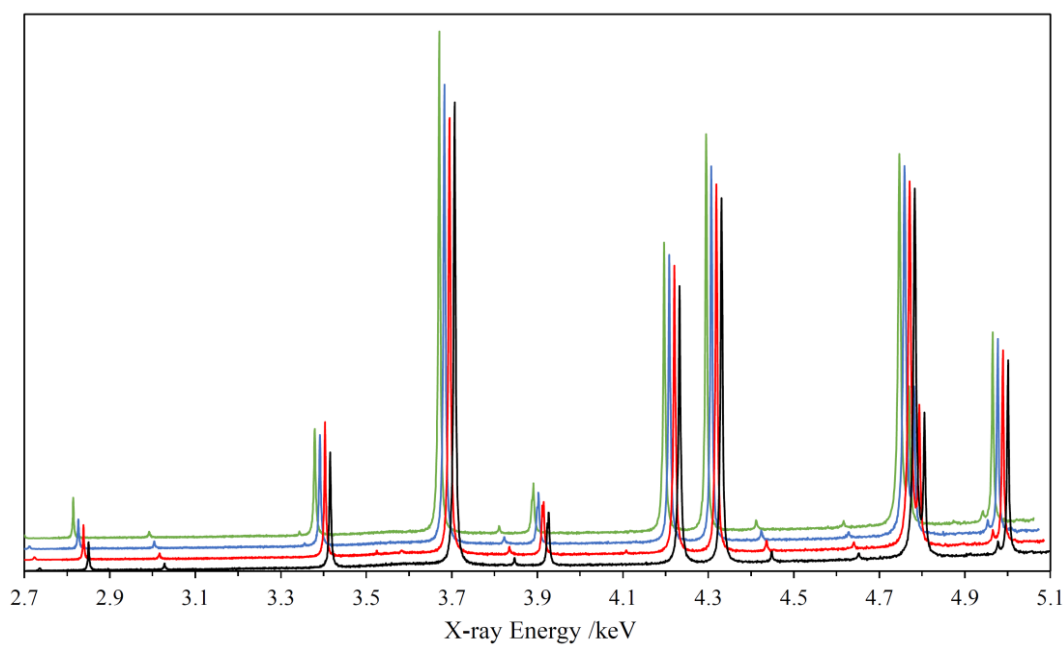

**Fig. S5.** EDXRD data for the Chrome Oxide Green paint illustrating relatively good powder averaging. The four detector channels have been offset relative to each other on both axes to allow easier comparison. Although the diffraction peak intensities are very similar in all four channels, some differences are apparent e.g. the most intense peak near 3.7 keV.

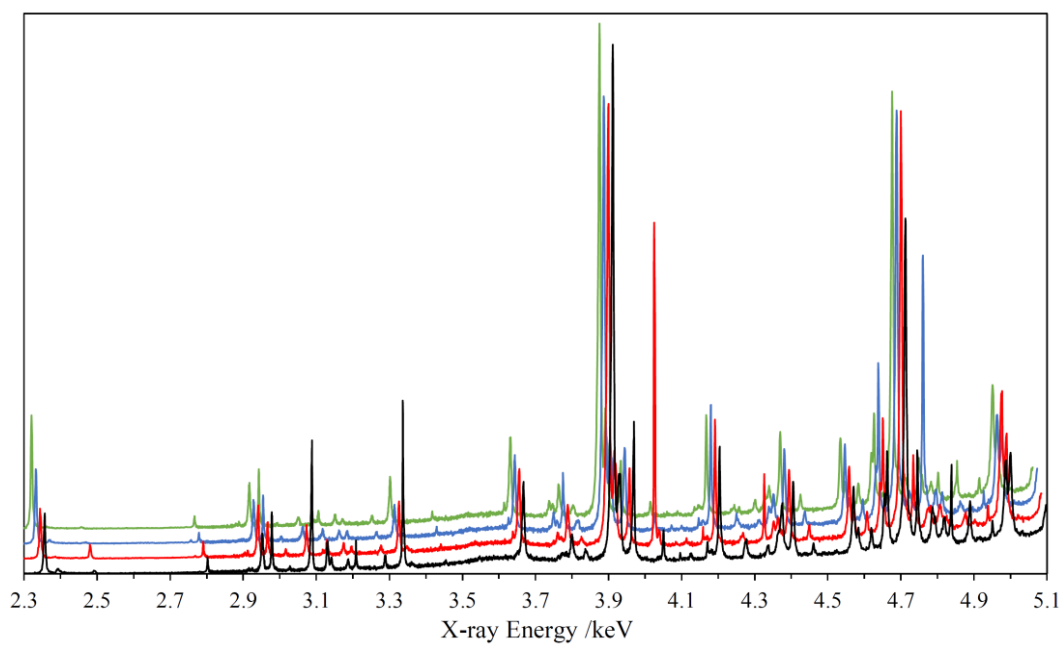

**Fig. S6.** EDXRD data for the Cremnitz White paint illustrating moderately good powder averaging ('intermediate behaviour'). The four detector channels have been offset relative to each other on both axes to allow easier comparison. Virtually all the diffraction peaks are present in all four channels, but with significant variations in intensity in some instances.

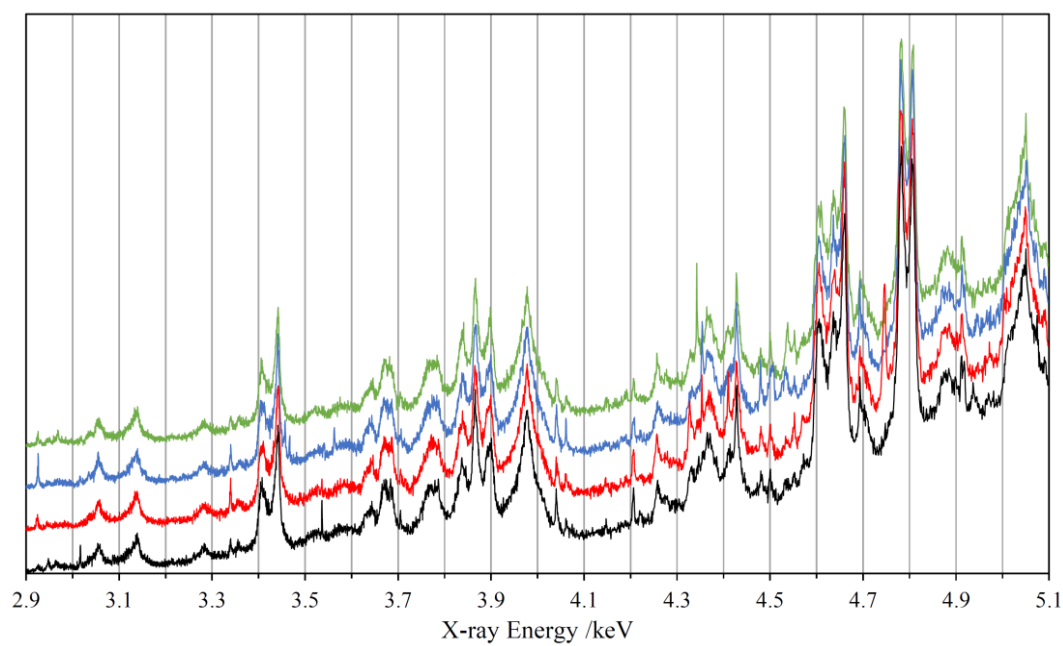

**Fig. S7.** EDXRD data for the Orange Molybdate paint illustrating variable powder averaging. The four detector channels have been offset relative to each other on the vertical axis to allow easier comparison. The broad diffraction peaks have very similar intensities in all four channels while there are a number of sharp peaks that appear on just one channel. The most obvious interpretation is the presence of (at least) two crystalline phases, one with small crystallites leading to good powder averaging and size-broadening of peaks, and the other with relatively large crystallites leading to poor powder averaging but very sharp peaks.

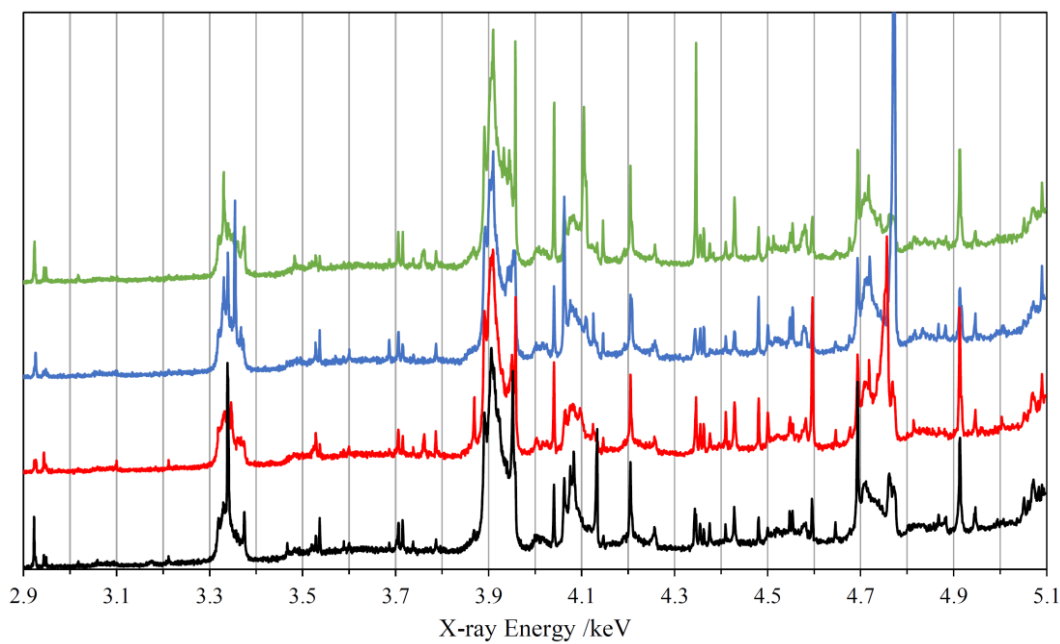

**Fig. S8.** EDXRD data for the Naples Yellow paint illustrating variable powder averaging. The four detector channels have been offset relative to each other on the vertical axis to allow easier comparison. Some diffraction peaks have very similar intensities in all four channels while others vary significantly.

## 5.    X-ray Penetration Depth

The penetration depth of X-rays into a material depends strongly on the X-ray energy (as well as the material composition and density). Fig. S9 shows the penetration depth of X-rays into a triglyceride (formula $C_{55}H_{98}O_6$, density 0.93 g cm$^{-3}$) representing an oil-based medium such as linseed oil. The depth shown is half of the 1/$e$ attenuation length in order to represent the depth at which X-ray photons can interact (diffract) and then re-emerge from the material with total 1/$e$ attenuation. It is important to note that the presence of other materials within the oil, such as pigments and fillers, will in most cases significantly reduce the penetration depth. The 1/$e$ attenuation length was calculated using the Center for X-ray Optics online calculator (https://henke.lbl.gov/optical_constants/).

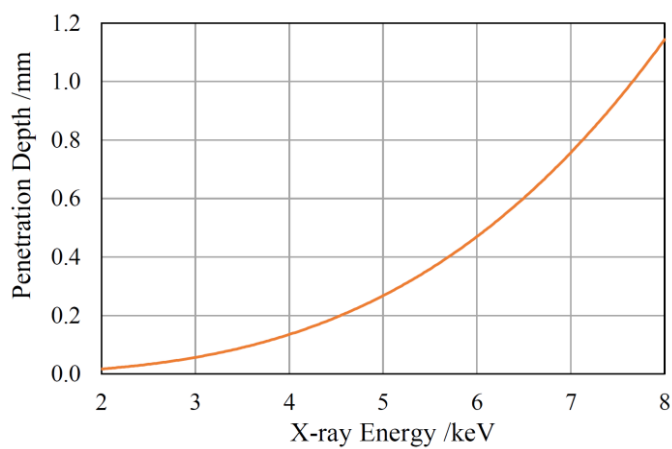

**Fig. S9.** Penetration depth of X-rays into a typical oil medium. See the main text for full details.

## 6. Comparison with Laboratory XRD

To place the performance of the back-reflection EDXRD technique in context, the results should ideally be compared with the results of another non-invasive XRD method using either portable or laboratory-based instrumentation. Unfortunately, the authors do not have access to an instrument of this type. Instead, data for two of the paints (Cobalt Blue and Cremnitz White) were acquired using a conventional laboratory XRD instrument. Small samples of the paints, on the order of a few mg, were removed from the test panel using a sharp scalpel. The Cremnitz White sample was ground to a powder by hand using a pestle and mortar but this method was ineffective for the Cobalt Blue sample which was finely diced with a razor blade instead. Each sample was sprinkled onto a low background silicon wafer and calibration was provided by the NIST Si standard (SRM 640a), also on a low background holder. In each case, the holder was spun at 15 rpm during data acquisition to improve powder averaging. Data was acquired with a Bruker D8 Advance using Cu-K$\alpha$ radiation and a Ni filter. The incident and diffracted beams were both collimated with 2.5° Soller slits and a Lynxeye XE position sensitive detector was used. Using the Si standard data, the instrumental FWHM resolution is 0.102° and 0.111° at 30° and 70°2$\theta$ respectively.

Direct comparisons of the laboratory and synchrotron data are shown in Figs S10 and S11 for Cobalt Blue and Cremnitz White respectively. The synchrotron EDXRD data has been converted to the angular equivalent using a wavelength of 1.5406 Å. There are small relative shifts in the diffraction peak positions because of sample height errors in the laboratory data. The synchrotron data has been scaled using a smooth, variable function to allow easier comparison of each observed diffraction peak, and the vertical axes are meaningful in terms of counts only for the laboratory data. The Si standard data is also shown in Fig. S10. The very sharp peaks seen in the synchrotron data is due to a small amount of corundum in the paint, unfortunately below the detection limit in the laboratory data. However, it is instructive to visually compare the synchrotron corundum peaks with the laboratory Si peaks as, in each case, the peak widths are dominated by the instrumental contribution with negligible sample contribution. The remaining peaks are due to the cobalt aluminate ($CoAl_2O_4$) pigment and are clearly broadened in both datasets.

Fig. S11 shows the same direct comparison for the Cremnitz White paint. Considerable differences in the relative peak intensities are apparent between the laboratory and synchrotron data. A Rietveld fit to the laboratory data shows that although the hydrocerussite phase shows preferred orientation (of the 00$l$ planes parallel to the sample surface), the strength of the effect is significantly smaller. As a consequence of the sample preparation method for the lab data, there is no reason to expect this parameter to remain the same and indeed this represents a loss of information relative to the undisturbed paint on the test panel. A second difference is that the Rietveld fit to the laboratory data shows a cerussite content of 31 wt% in the two-phase mixture, compared to 19 wt% derived from the synchrotron data. The powder averaging is significantly better in the laboratory data and it is likely that the poorer powder averaging adversely affects the quantification accuracy of the synchrotron data. The resolution of the laboratory data was insufficient to support an analysis of the anisotropic crystallite shape model fitted to the synchrotron data.

It is not surprising that the quality of the synchrotron data, particularly regarding instrumental resolution and signal-to-noise ratios, is superior. However, it is emphasised that the authors plan to implement the back-reflection EDXRD technique in the laboratory using microcalorimeter technology which is expected to lead to only a modest loss of resolution and comparable signal-to-noise for the same acquisition times (because of the broadband acquisition advantage). The synchrotron data is therefore representative of the expected performance of laboratory back-reflection EDXRD.

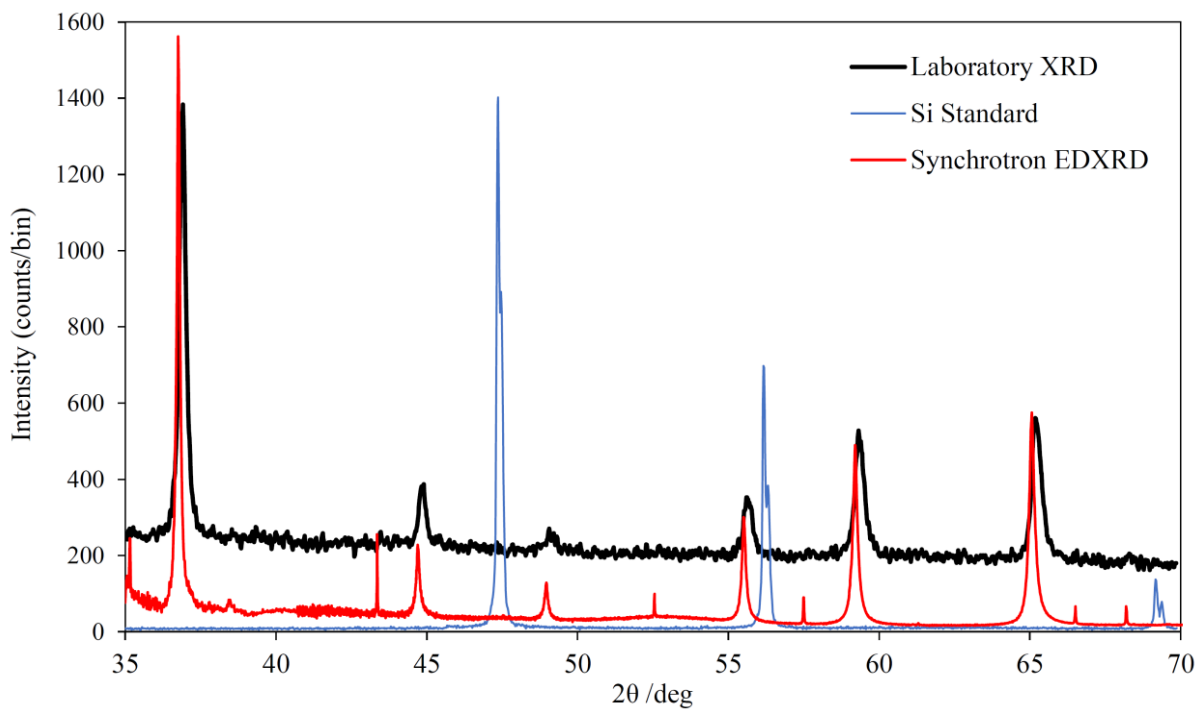

**Fig. S10.** Comparison of laboratory and synchrotron XRD data for the Cobalt Blue paint. The NIST Si standard laboratory data is also shown (scaled by a factor of 0.2 relative to the Cobalt Blue pattern).

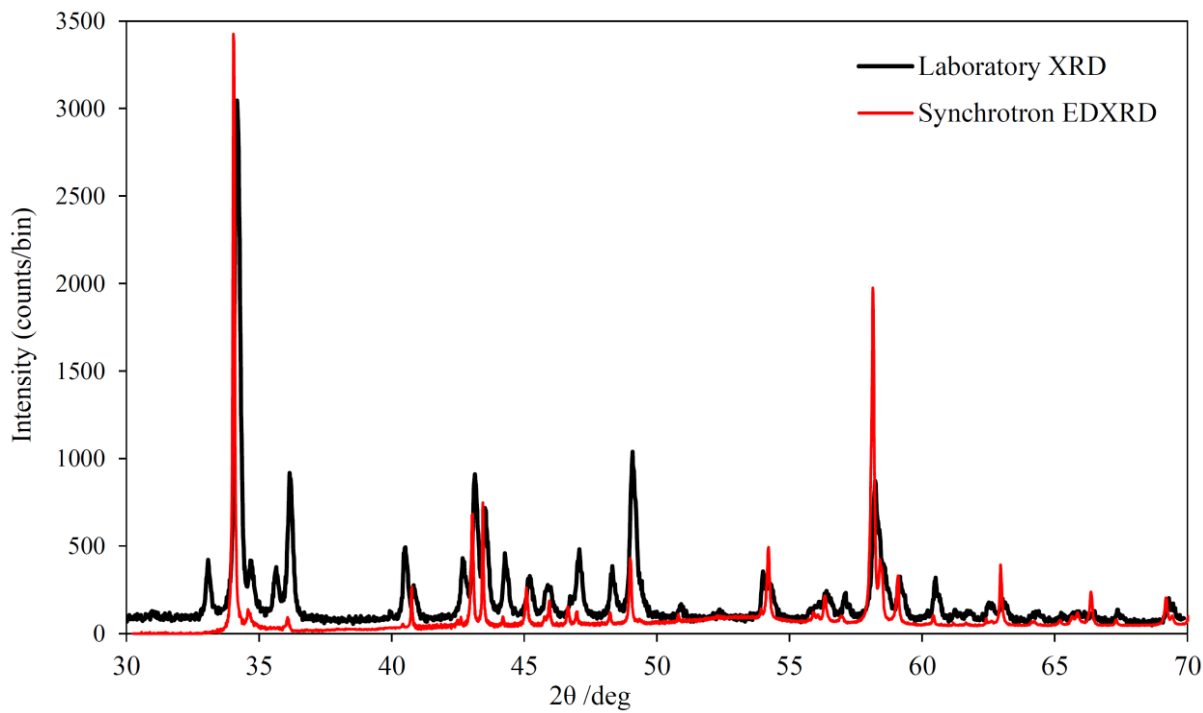

**Fig. S11.** Comparison of laboratory and synchrotron XRD data for the Cremnitz White paint.

## 7. Artists' Paints

This section reports the results for the paints not described in the Results section of the main article. The paints have been categorised into four groups A – D, defined in Section 3 of the main article (Table S3). The diffraction patterns are shown along with Pawley fits and Rietveld refinements if an analysis was possible. For all of these paints, the diffraction peaks are broadened by sample effects i.e. the peaks are broader than the observed diffraction peaks of the $LaB_6$ standard for which the broadening is instrumental (primarily geometrical). In most cases no attempt has been made to distinguish between the most common causes of peak broadening – small crystallite mean size and microstrain. Except where otherwise noted, size broadening effects have been assumed whenever the broadening appears to be isotropic (with no *hkl* dependence). The reported crystallite domain sizes should be interpreted in this context. The same approach has been taken even for phases that exhibit poor powder averaging despite the fact that this observation implies the presence of larger crystallites.

For some of the diffraction patterns, absorption edge jumps and the associated EXAFS pattern have been subtracted from the data using a simple algorithm that takes account of the overlap between the 'moving window' [1] and the Kβ fluorescence peak corresponding to the element responsible for the absorption edge. The intensity of the Kβ peak at each monochromator energy was estimated by scaling the observed Kα peak intensity (the Kα peak is fully resolved from the scattered/diffracted monochromator energy by the SDDs, unlike the Kβ peak). This method has been implemented only for K-edges.

**Table S3.** Panel paints categorised into groups A – D.

| Group A | | Group B | Group C | Group D |
|---|---|---|---|---|
| Cerulean Blue | Minium | Manganese Blue | Terre Vert | Phthalo Blue Lake |
| Cobalt Blue | Cadmium Red | Dioxazine Violet | Orange Molybdate | Phthalo Green Lake |
| Ultramarine Blue | French Yellow Ochre | Bright Yellow Lake | Viridian | Yellow Lake |
| Manganese Violet | Raw Sienna | Pyrrolo Vermilion | | Naphthol Red |
| Chrome Oxide Green | Raw Umber | | | Scarlet Lake |
| Cobalt Turquoise | Transparent Red Oxide | | | Alizarin Crimson |
| Chrome Green | Cremnitz White | | | Magenta |
| Lemon Yellow | Flake White | | | Lamp Black |
| Aureolin | Zinc White | | | |
| Naples Yellow | Flemish White | | | |
| Cadmium Gold Yellow | Titanium White No. 3 | | | |
| Chrome Yellow | Ivory Black[a] | | | |
| Vermilion | | | | |

[a]The black pigment in this paint is expected to be carbonaceous material which has not been detected (detection of the pigment is the definition for group A paints). However, the pigmentation is derived from charred animal bones and hydroxyapatite has been detected; this mineral is intimately associated with the pigment, justifying categorisation in group A.

### 6.1 Group A Paints

**Cerulean Blue (Michael Harding No. 603)**

The strongest diffraction peaks were readily fitted by those of a spinel with all remaining peaks attributed to $Al_2O_3$ and $BaSO_4$ (Fig. S10a). The spinel is presumed to have nominal formula $Co(Al_{1-x}Cr_x)_2O_4$, based on previous elemental and Raman analysis of the paint [2]. A Rietveld refinement was carried out to determine phase fractions (Fig. S10b, Table S4); the two detectors exhibited different intensities for the $BaSO_4$ peaks and so phase quantification error is likely larger than the estimated standard uncertainty quoted. In order to determine the extent of Cr substitution in the spinel phase from the XRD pattern, two analytical approaches can be taken:

**Table S4.** Refined crystallographic parameters for the identified phases in Cerulean Blue.

| | Literature | Pawley Fit | Rietveld Fit |
|---|---|---|---|
| **Co(Al$_{1-x}$Cr$_x$)$_2$O$_4$ ($Fd\bar{3}m$)** | ICSD Code 78402[a] | | |
| $a$ (Å) | 8.10664 (4) | 8.18999 (5) | 8.18974 (9) |
| Crystallite size (nm) | | 36.3 (2) | 35.2 (4) |
| $x$ | | 0.37[b] | 0.214 (15) |
| Weight fraction (%) | | | 82.2 (6) |
| **Al$_2$O$_3$ ($R\bar{3}c$)** | ICSD Code 51687 | | |
| $a$ (Å) | 4.7597 (1) | 4.7779 (2) | 4.7774 (3) |
| $c$ (Å) | 12.9935 (3) | 13.0440 (8) | 13.0385 (16) |
| Crystallite size (nm) | | 37.3 (12) | 45 (3) |
| Weight fraction (%) | | | 10.3 (5) |
| **BaSO$_4$ (*Pnma*)** | ICSD Code 200112 | | |
| $a$ (Å) | 8.8842 (12) | 8.87491 (7) | 8.87465 (19) |
| $b$ (Å) | 5.4559 (8) | 5.45321 (5) | 5.45351 (9) |
| $c$ (Å) | 7.1569 (9) | 7.15351 (5) | 7.15341 (16) |
| Crystallite size (nm) | | 222 (4) | 247 (13) |
| Weight fraction (%) | | | 7.6[c] (3) |
| $R_{wp}$ (%) | | 6.08 | 14.8 |

[a]CoAl$_2$O$_4$.
[b]Inferred from Vegard's Law, see text.
[c]This phase has poor powder averaging.

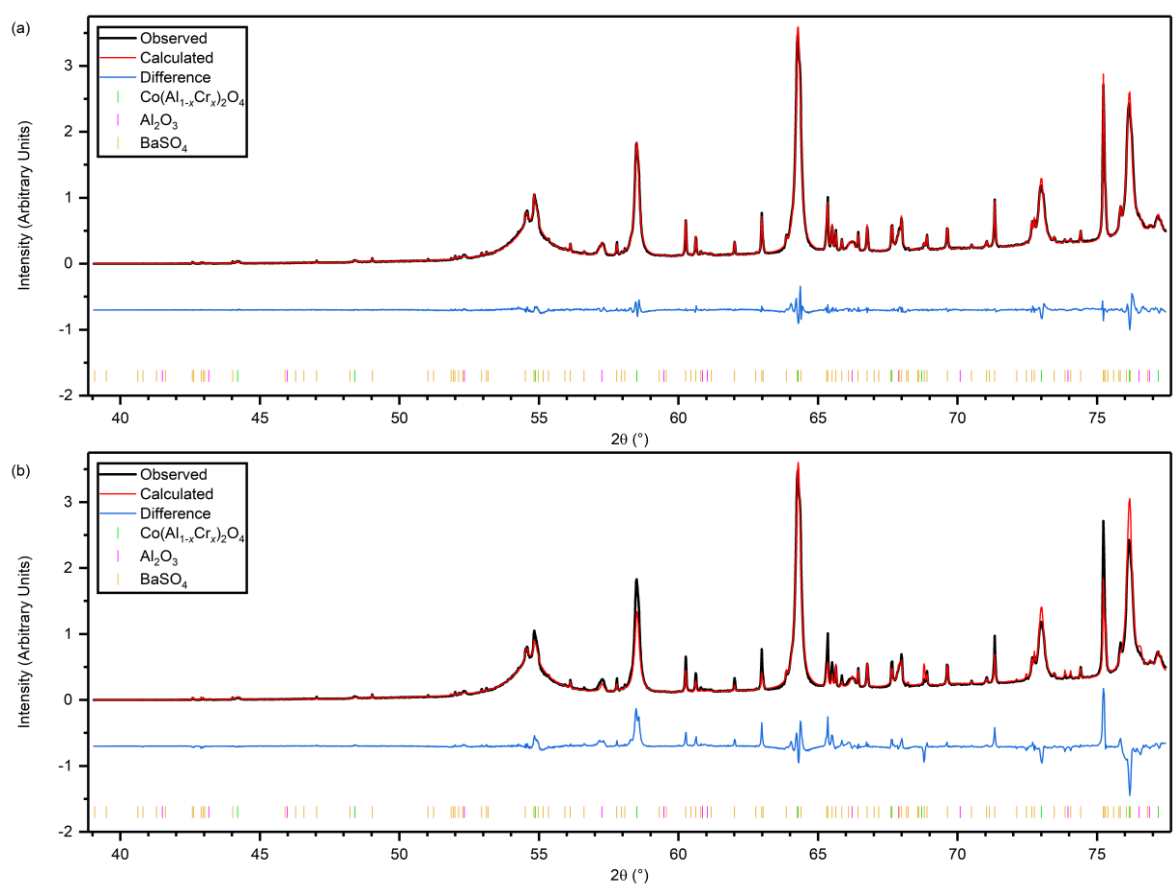

**Fig. S12.** (a) Pawley and (b) Rietveld fits to the Cerulean Blue diffraction pattern. In each case the difference plot is shown offset on the vertical scale. The broad peak centred on $2\theta = 54.8°$ is due to diffraction by the graphitic vacuum window, and was fitted by inclusion of two pseudo-Voigt functions.

1. *Vegard's Law*

The value of $x$ can be inferred from the lattice parameter using Vegard's law. Only the end-members, $CoAl_2O_4$ and $CoCr_2O_4$, are available on the ICSD (Codes 78402 and 61612, respectively) with lattice parameters 8.10664 (4) Å and 8.3346 (3) Å, respectively. Based on the lattice parameter extracted from the Pawley fit of 8.18999 (5) Å (Table S4), $x$ can be interpolated as 0.37 (1) (the error estimate is based only on prior experience of accuracy using lattice parameter interpolation).

2. *Site Occupancy Refinement*

As part of the Rietveld fit, the relative occupancy of the (Al,Cr) 16$d$ site can be refined. The total occupancy was constrained to unity and $x$ refined to 0.214 (15). All other crystallographic parameters (O coordinates, atomic displacement parameters) remained fixed at the values obtained from the ICSD Code 78402 crystal structure.

The values of $x$ obtained from the two approaches are significantly different and both have assumptions that limit the reliability of these analyses. The former assumes that the solid solution obeys Vegard's law (i.e. linearity of lattice parameters vs $x$), whilst the latter assumes that there is no preferred orientation of the crystallites. Both approaches assume that Cr is the only significant substituent in the spinel and that the crystal structure is otherwise unchanged. On the basis that Vegard's law is accurately obeyed for related spinels [e.g. $Zn(Al_{1-x}Cr_x)_2O_4$, see ref. 3], it is believed that the estimate of $x$ based on lattice parameter interpolation is more reliable.

It is noted that the refined unit cell parameters of $Al_2O_3$ show a slightly enlarged cell, possibly due to Co or Cr substitution into the corundum structure, but no attempt was made to quantify this.

**Cobalt Blue (Michael Harding No. 506)**

Table S5. Refined crystallographic parameters for the identified phases in Cobalt Blue.

|  | Literature | Pawley Fit | Rietveld Fit |
|---|---|---|---|
| **$CoAl_2O_4$ ($Fd\bar{3}m$)** | ICSD Code 78402 | | |
| $a$ (Å) | 8.10664 (4) | 8.10236 (2) | 8.10238 (2) |
| Crystallite size (nm) | | 33.14 (8) | 33.65 (11) |
| Weight fraction (%) | | | 96.91 (10) |
| **$Al_2O_3$ ($R\bar{3}c$)** | ICSD Code 51687 | | |
| $a$ (Å) | 4.7597 (1) | 4.75945 (3) | 4.75944 (4) |
| $c$ (Å) | 12.9935 (3) | 12.99219 (12) | 12.99213 (18) |
| Crystallite size (nm) | | 291 (13) | 280 (20) |
| Weight fraction (%) | | | 3.09 (10) |
| $R_{wp}$ (%) | | 6.16 | 8.32 |

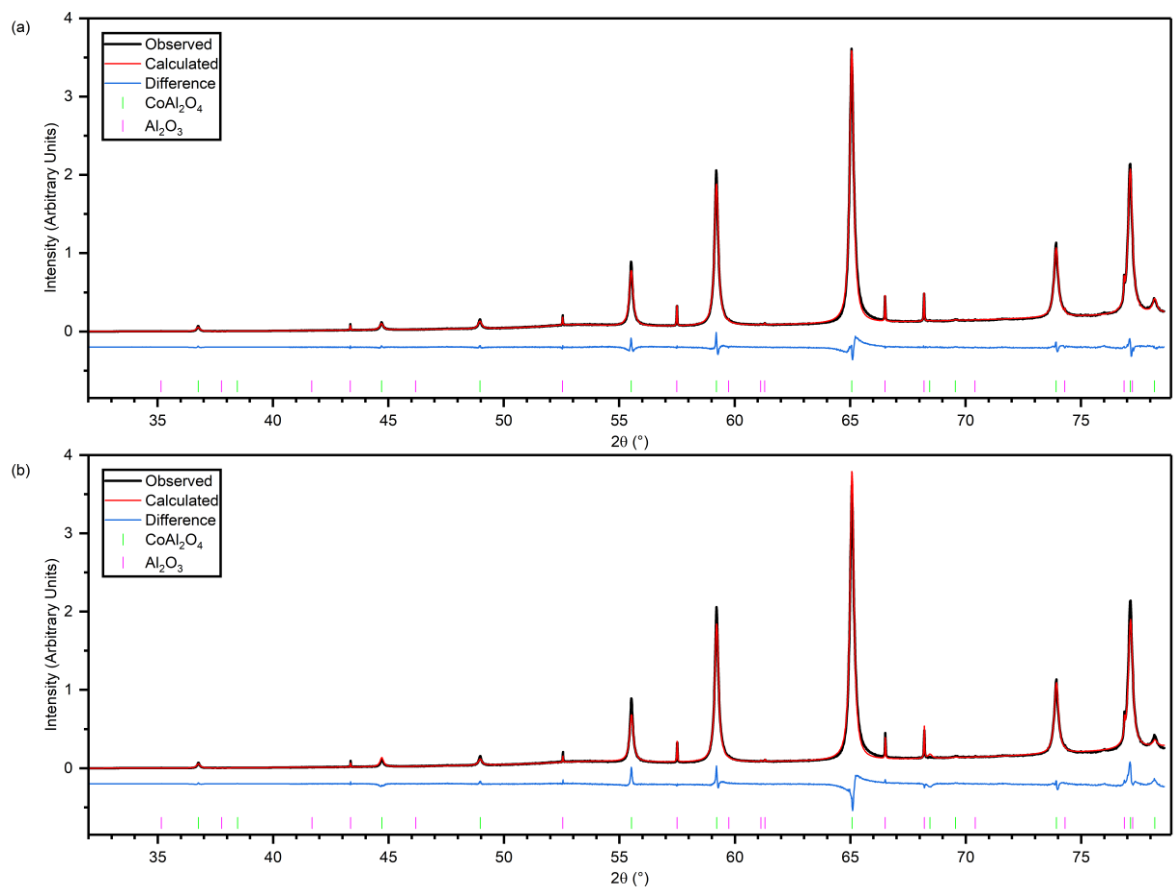

**Fig. S13.** (a) Pawley and (b) Rietveld fits to the Cobalt Blue diffraction pattern. In each case the difference plot is shown offset on the vertical scale.

Similar to Cerulean Blue, Cobalt Blue contains a spinel and weak diffraction peaks attributed to $Al_2O_3$. The lattice parameter of the spinel is consistent with that of $CoAl_2O_4$ (Table S5). A Rietveld refinement (Fig. S13b) was carried out in order to determine phase fractions.

**Ultramarine Blue (Michael Harding No. 113)**

Table S6. Refined crystallographic parameters for the identified phase in Ultramarine Blue.

|  | Literature | Pawley Fit |
|---|---|---|
| **Lazurite ($P\bar{4}3n$)** | ICSD Code 49759[a] | |
| $a$ (Å) | 9.105 (2) | 9.07944 (12) |
| Crystallite size (nm) |  | 25.5 (3) |
| $R_{wp}$ (%) |  | 5.60 |

[a]Formula $Na_{8.56}(Al_6Si_6O_{24})(SO_4)_{1.56}S_{0.44}$.

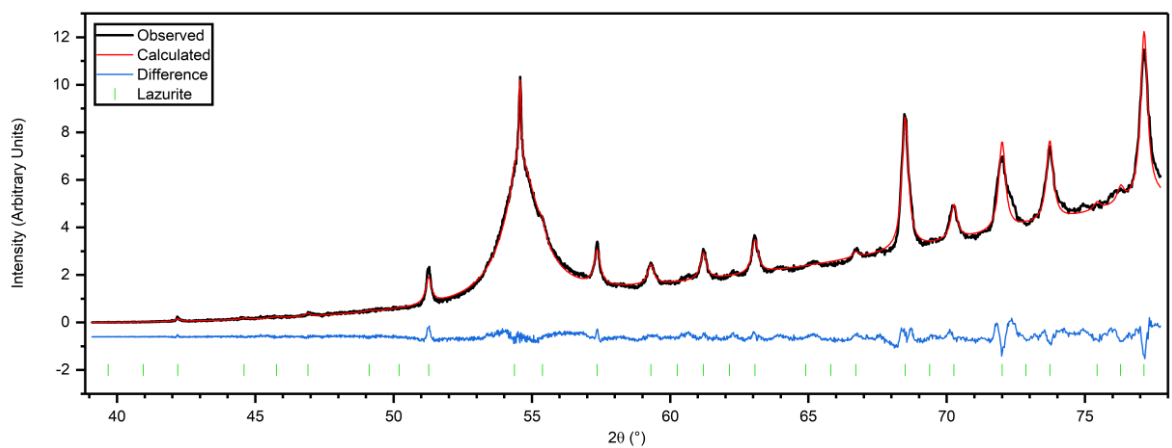

**Fig. S14.** Pawley fit to the Ultramarine Blue diffraction pattern. The difference plot is shown offset on the vertical scale. The broad peak centred on $2\theta = 54.8°$ is due to diffraction by the graphitic vacuum window, and was fitted by inclusion of two pseudo-Voigt functions.

The diffraction pattern is consistent with the presence of a sodalite framework structure adopted by lazurite (Fig. S14, Table S6). The structure-type accommodates a wide range of compositional variety and the data were insufficient to allow a full structural model to be constructed.

# Manganese Violet (Michael Harding No. 304)

**Table S7**. Refined crystallographic parameters for the identified phases in Manganese Violet.

|  | Literature | Pawley Fit | Rietveld Fit |
| --- | --- | --- | --- |
| **α-NH$_4$MnP$_2$O$_7$ (*P*2$_1$/*c*)** | Begum and Wright [4] | | |
| *a* (Å) | 7.4252 (3) | 7.4150 (2) | 7.4147 (3) |
| *b* (Å) | 9.6990 (4) | 9.7047 (3) | 9.7049 (3) |
| *c* (Å) | 8.6552 (4) | 8.6451 (3) | 8.6457 (3) |
| *β* (°) | 105.627 (3) | 105.551 (2) | 105.545 (3) |
| Crystallite size (nm) | | 102 (3) | 111 (5) |
| Weight fraction (%) | 69 (3) | | 71 (2) |
| **β-NH$_4$MnP$_2$O$_7$ (*P$\bar{1}$*)** | Begum and Wright [4] | | |
| *a* (Å) | 8.4034 (6) | 8.40359 (14) | 8.4036 (2) |
| *b* (Å) | 6.1498 (4) | 6.14997 (5) | 6.1497 (2) |
| *c* (Å) | 6.1071 (4) | 6.10641 (6) | 6.1064 (2) |
| *α* (°) | 104.618 (5) | 104.6288 (11) | 104.620 (3) |
| *β* (°) | 100.748 (5) | 100.7411 (10) | 100.743 (3) |
| *γ* (°) | 96.802 (6) | 96.8044 (11) | 96.808 (3) |
| Crystallite size (nm) | | 480 (40) | 300 (35) |
| Weight fraction (%) | 31 (3) | | 29[a] (2) |
| *R$_{wp}$* (%) | | 11.3 | 33.1 |

[a]This phase has poor powder averaging.

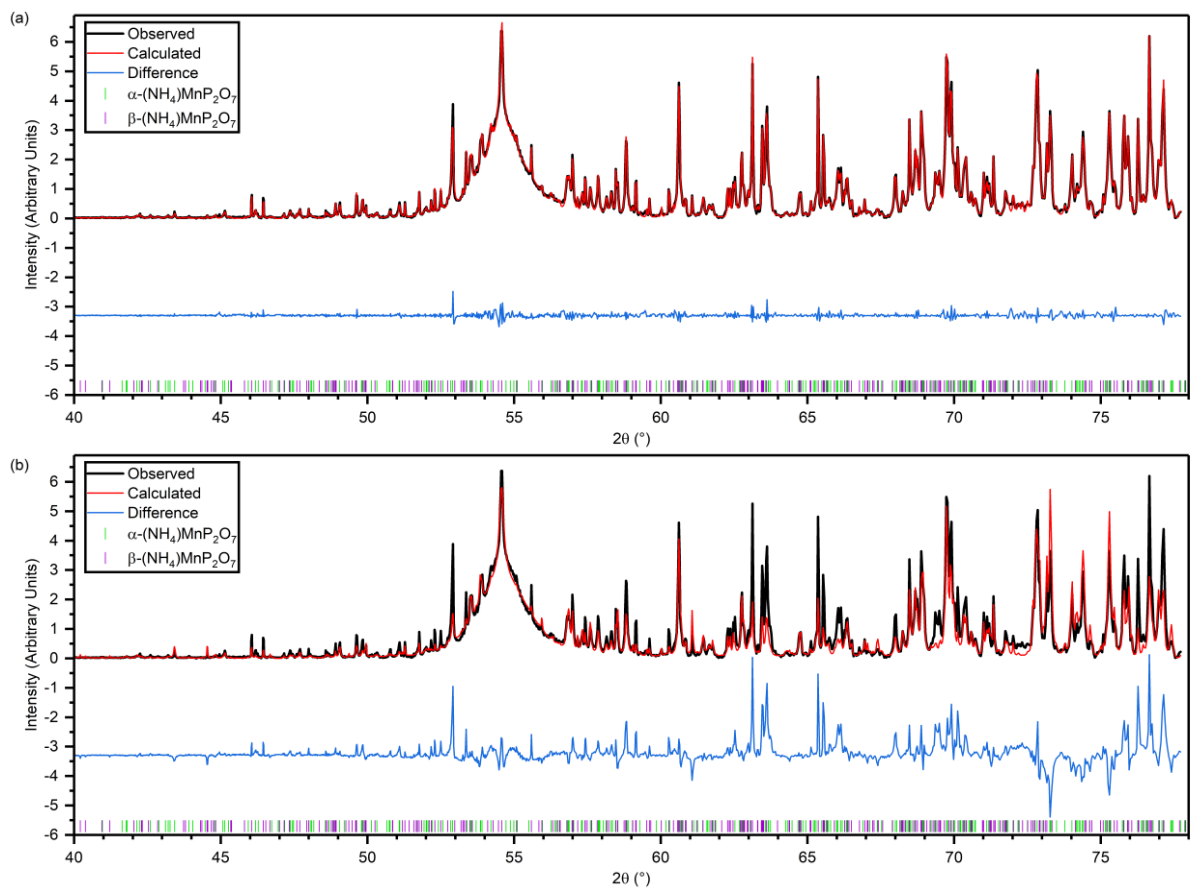

**Fig. S15.** (a) Pawley and (b) Rietveld fits to the Manganese Violet diffraction pattern. In each case the difference plot is shown offset on the vertical scale. The broad peak centred on 2*θ* = 54.8° is due to diffraction by the graphitic vacuum window, and was fitted by inclusion of two pseudo-Voigt functions.

The Rietveld refinement provides a significantly poorer fit than the Pawley fit (Fig. S15), demonstrated by the larger $R_{wp}$ value (Table S7). This is mostly due to poor powder averaging of the β-$NH_4MnP_2O_7$ phase which is likely to reduce the accuracy of the phase quantification but it is assumed that the large number of peaks aids in 'averaging out' any errors. The analysis of this paint is described in more detail in the main article.

**Chrome Oxide Green (Michael Harding No. 305)**

Table S8. Refined crystallographic parameters for the identified phase in Chrome Oxide Green.

|  | Literature | Pawley Fit |
|---|---|---|
| **$Cr_2O_3$ ($R\bar{3}c$)** | ICSD Code 75577 |  |
| a (Å) | 4.9570 (3) | 4.95905 (1) |
| c (Å) | 13.5923 (2) | 13.59696 (3) |
| Crystallite size (nm) |  | 70.6 (2) |
| $R_{wp}$ (%) |  | 8.11 |

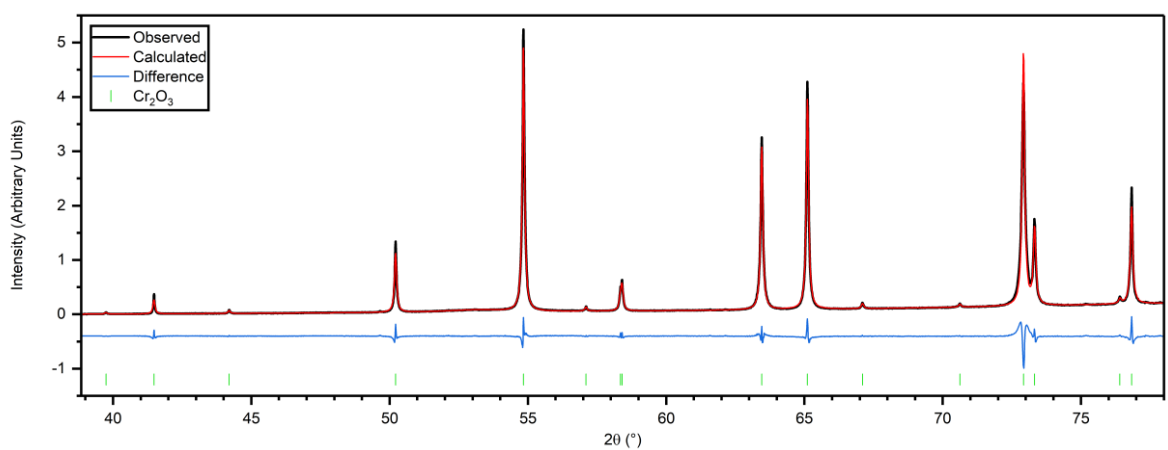

**Fig. S16.** Pawley fit to the Chrome Oxide Green diffraction pattern. The difference plot is shown offset on the vertical scale.

All diffraction peaks can be indexed to a corundum-type cell (Fig. S16) consistent with $Cr_2O_3$, with a mean crystallite size of 70.6 (2) nm (Table S8).

**Cobalt Turquoise (Michael Harding No. 507)**

Table 9. Refined crystallographic parameters for the identified phases in Cobalt Turquoise.

|  | Literature | Pawley Fit[a] |  |
|---|---|---|---|
| **$CoCr_2O_4$ ($Fd\bar{3}m$)** | ICSD Code 61612 |  |  |
| a (Å) | 8.3346 (3) | 8.32044 (1) | 8.32992 (3) |
| Crystallite size (nm) |  | 72.5 (2) |  |
| **$Cr_2O_3$ ($R\bar{3}c$)** | ICSD Code 75577 |  |  |
| a (Å) | 4.9570 (3) | 4.9522 (2) | 4.95764 (15) |
| c (Å) | 13.5923 (2) | 13.5663 (7) | 13.5910 (7) |
| Crystallite size (nm) |  | 79 (3) |  |
| $R_{wp}$ (%) |  | 4.77 |  |

[a]Two distinct compositions were observed for each phase.

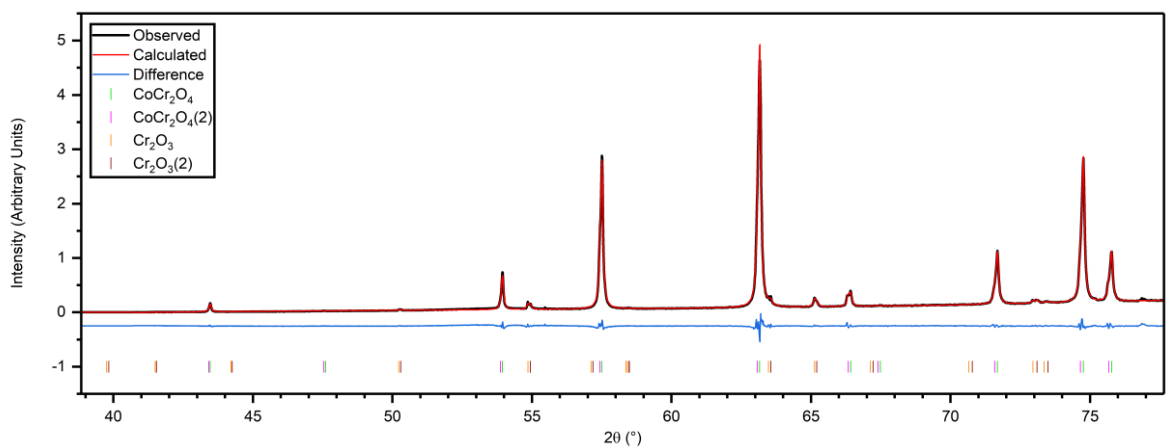

**Fig. S17.** Pawley fit to the Cobalt Turquoise diffraction pattern. The difference plot is shown offset on the vertical scale.

The diffraction pattern of this paint contains peaks attributable to the spinel $CoCr_2O_4$ and to $Cr_2O_3$ (Fig. S17). However, close study of the pattern reveals splitting in the peaks of both phases consistent with two distinct compositions for each phase. In order to fit this robustly, the primary and secondary compositions were constrained to have the same fitted widths (and therefore identical derived mean crystallite sizes). In both cases the second composition has cell parameters 0.1 – 0.2% smaller, indicative of a small degree of substitution by an unidentified element, most likely to be either Co or Al. It is likely that the $Cr_2O_3$ phase represents left-over raw material from the synthesis of the spinel and therefore that the presence of two distinct spinel phases is a direct consequence of the presence of two $Cr_2O_3$ phases.

Observation of the peak splitting was possible only because of the very high resolution of the technique. It may be that the bimodal characteristic of these phases is restricted to one supplier and perhaps for a limited period of time, offering an additional potential method to distinguish otherwise similar paints for the purposes of authentication or art history research. Similar peak splitting has also been observed for the crystalline phases in Lemon Yellow and Aureolin.

**Chrome Green (Rublev Colours)**

**Table S10.** Refined crystallographic parameters for the identified phase in Chrome Green.

|  | Literature | Pawley Fit |
|---|---|---|
| **$PbCr_{1-x}S_xO_4$ (*Pnma*)** | Monico *et al.* [5][a] | |
| *a* (Å) | 8.592 (2) | 8.585 (3) |
| *b* (Å) | 5.527 (2) | 5.514 (2) |
| *c* (Å) | 7.061 (2) | 7.069 (2) |
| Crystallite size (nm) |  | 8.5 (7) |
| $R_{wp}$ (%) |  | 6.59 |

[a] $x = 0.2$.

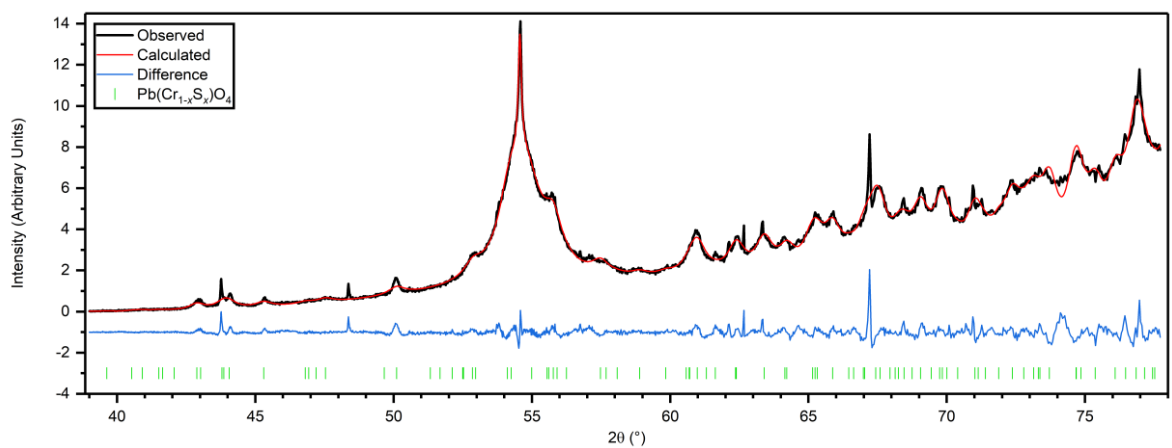

**Fig. S18.** Pawley fit to the Chrome Green diffraction pattern. The difference plot is shown offset on the vertical scale. The broad peak centred on $2\theta = 54.8°$ is due to diffraction by the graphitic vacuum window, and was fitted by inclusion of two pseudo-Voigt functions.

According to the supplier, the Chrome Green paint contains a mixture of the pigments chrome yellow (lead chromate, $PbCrO_4$) and Prussian blue (ferric ferrocyanide, $Fe_4[Fe(CN)_6]_3 \cdot xH_2O$). The Chrome Green diffraction pattern can be fitted with an orthorhombic phase corresponding to $PbCr_{1-x}S_xO_4$. The fitted unit cell dimensions are close to those reported by Monico *et al* [5] for $x = 0.2$, though slightly smaller which indicates marginally greater S content (Table S10). $PbCr_{1-x}S_xO_4$ occurs in both monoclinic and orthorhombic forms; the latter has a miscibility gap from $x = 0.2$ to 0.9 and is metastable with respect to the monoclinic phase for low values of $x$ [6].

Some unidentified sharp diffraction peaks with poor powder averaging can be seen at $2\theta = 43.8°$, $48.4°$, $62.7°$, $67.2°$, $70.9°$ and $77.0°$ (Fig. S18). These peaks do not correspond to the positions expected for Prussian blue.

**Lemon Yellow (Michael Harding No. 108)**

Table S11. Refined crystallographic parameters for the identified phase in Lemon Yellow.

|  | Literature | Pawley Fit |
|---|---|---|
| **$BaCrO_4$ (*Pnma*)** | ICSD Code 62560 |  |
| $a$ (Å) | 9.113 (4) | 9.11019 (2) |
| $b$ (Å) | 5.528 (3) | 5.53102 (2) |
| $c$ (Å) | 7.336 (4) | 7.33603 (2) |
| Crystallite size (nm) |  | 159.5 (8) |
| $R_{wp}$ (%) |  | 9.66 |

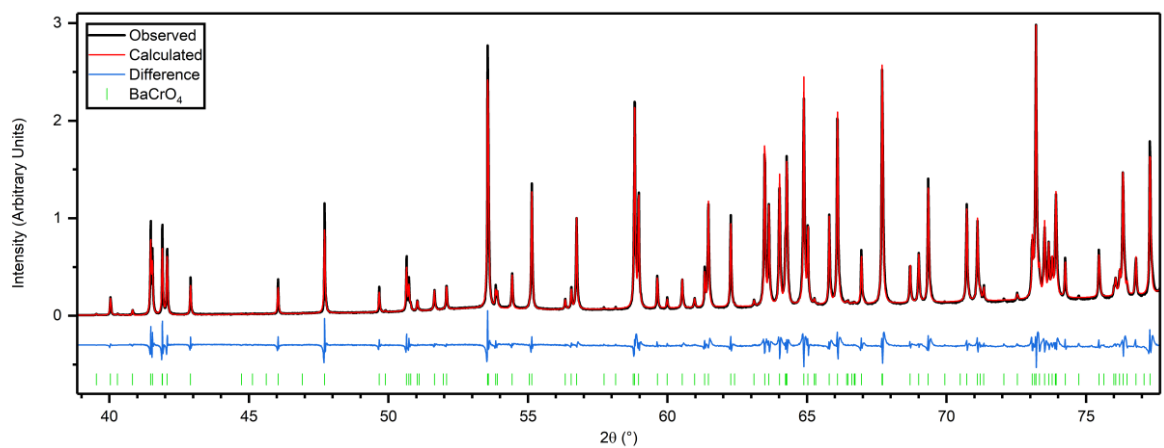

**Fig. S19.** Pawley fit to the Lemon Yellow diffraction pattern. The difference plot is shown offset on the vertical scale.

The only crystalline phase observed in Lemon Yellow was $BaCrO_4$ (Table S11). A small shoulder on the high angle side of each diffraction peak was noted (Fig. S19). This peak asymmetry is likely to be caused by a second phase of slightly different composition.

### Aureolin (Michael Harding No. 501)

**Table S12.** Refined crystallographic parameters for the identified phases in Aureolin.

|  | Literature | Pawley Fit[b] |  |
|---|---|---|---|
| $K_3[Co(NO_2)_6]$ ($Fm\bar{3}$) | Vendilo et al. [7][a] |  |  |
| $a$ (Å) | 10.468 (6) – 10.498 (7) | 10.4861 (5) | 10.4953 (4) |
| Crystallite size (nm) |  | 61 (2) |  |
| $R_{wp}$ (%) |  | 6.71 |  |

[a]Multiple compositions reported.
[b]Two distinct compositions observed.

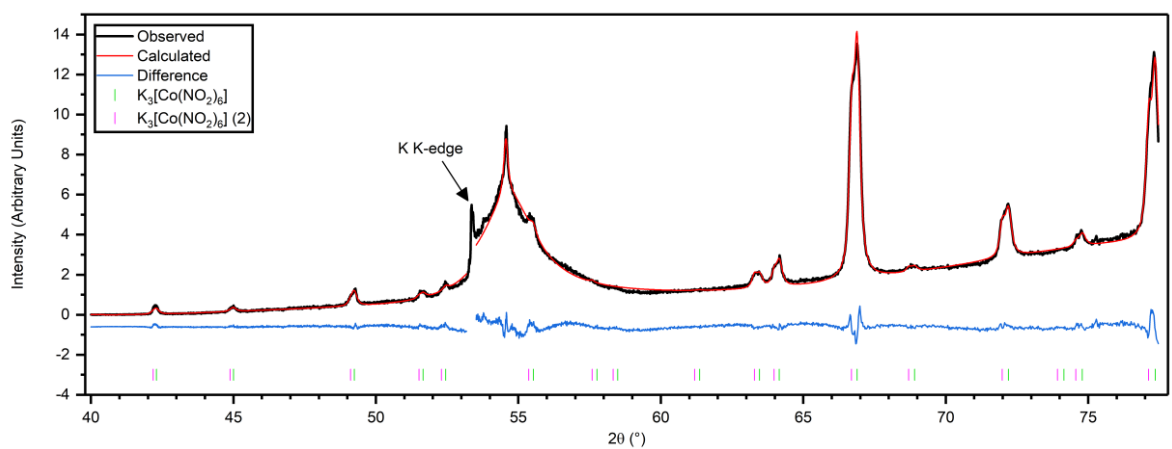

**Fig. S20.** Pawley fit to the Aureolin diffraction pattern. The difference plot is shown offset on the vertical scale. The broad peak centred on $2\theta = 54.8°$ is due to diffraction by the graphitic vacuum window, and was fitted by inclusion of two pseudo-Voigt functions. The diffraction pattern included a strong K K-edge at $2\theta = 53.3°$ that was difficult to remove due to its proximity to the vacuum window diffraction feature. This region was excluded from the fit.

Aureolin can be fitted by two $K_3[Co(NO_2)_6]$ phases with lattice parameters 10.4953 (4) Å and 10.4861 (5) Å (Table S12). The lattice parameters fall within the range reported by Vendilo et al. [7] who suggested that varying amounts of Na and $H_2O$ were responsible for the differing lattice parameters. Due to the heavy overlap of peaks the crystallite size of both phases was constrained to be the same.

**Chrome Yellow (Rublev Colours)**

Table S13. Refined crystallographic parameters for the identified phases in Chrome Yellow.

|  | Literature | Pawley Fit | Rietveld Fit |
|---|---|---|---|
| **PbCrO$_4$ ($P2_1/n$)** | ICSD Code 40920 | | |
| $a$ (Å) | 7.127 (2) | 7.1216 (2) | 7.1233 (3) |
| $b$ (Å) | 7.438 (2) | 7.4335 (2) | 7.4336 (3) |
| $c$ (Å) | 6.799 (2) | 6.7996 (2) | 6.8006 (2) |
| $\beta$ (°) | 102.43 (2) | 102.434 (3) | 102.448 (4) |
| Crystallite size (nm) | | 64.0 (11) | 69 (2) |
| Weight fraction (%) | | | 94 (1) |
| **BaSO$_4$ ($Pnma$)** | ICSD Code 200112 | | |
| $a$ (Å) | 8.8842 (12) | 8.8691 (6) | 8.873 (3) |
| $b$ (Å) | 5.4559 (8) | 5.4511 (5) | 5.455 (2) |
| $c$ (Å) | 7.1569 (9) | 7.1504 (4) | 7.152 (3) |
| Crystallite size (nm) | | 90 (6) | 64 (19) |
| Weight fraction (%) | | | 6 (1) |
| $R_{wp}$ (%) | | 8.19 | 17.8 |

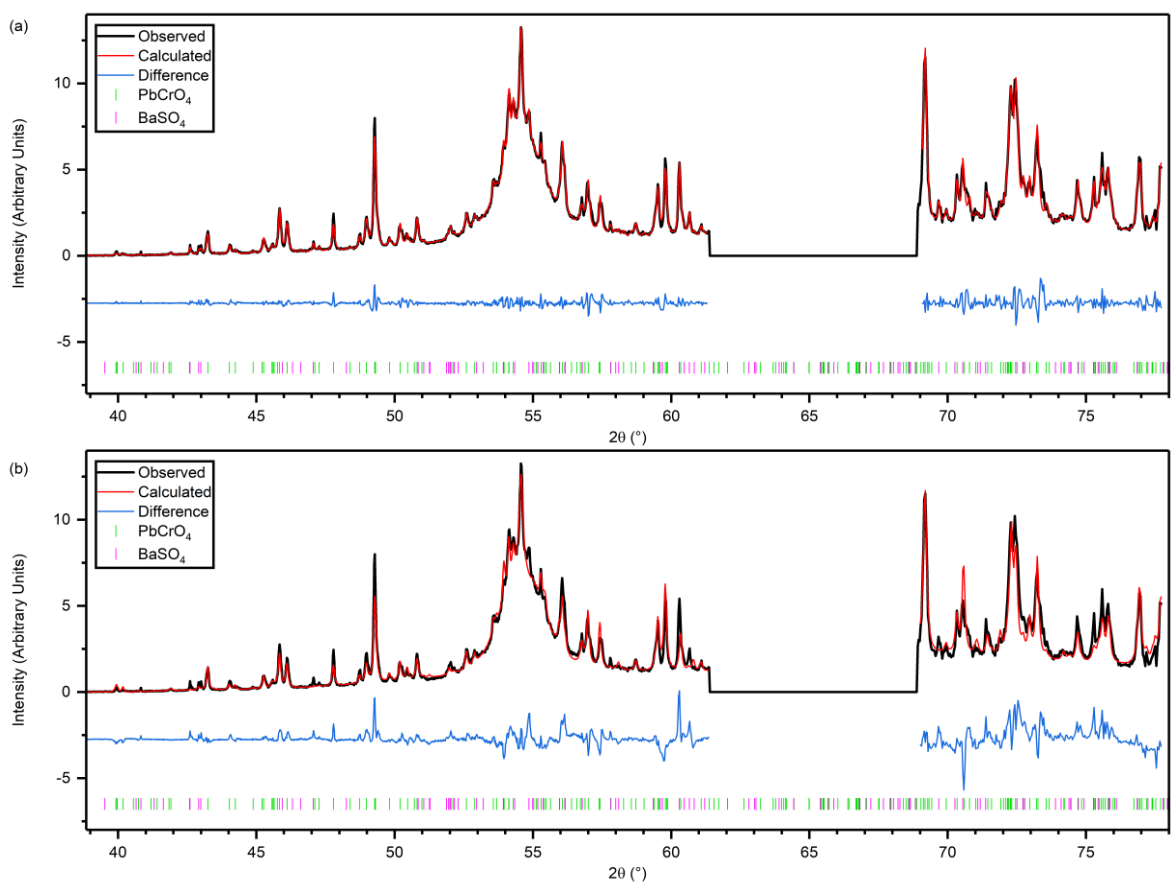

**Fig. S21.** (a) Pawley and (b) Rietveld fits to the Chrome Yellow diffraction pattern. In each case the difference plot is shown offset on the vertical scale. The broad peak centred on $2\theta = 54.8°$ is due to diffraction by the graphitic vacuum window, and was fitted by inclusion of two pseudo-Voigt functions.

A data acquisition error meant that no data was collected in the range 61.4° < 2θ < 68.9° (Fig. S21). The phases present, PbCrO$_4$ and BaSO$_4$ (Table S13), could be readily identified despite the missing data.

**Vermilion (Rublev Colours)**

Table S14. Refined crystallographic parameters for the identified phase in Vermilion.

|  | Literature | Pawley Fit |
|---|---|---|
| **HgS (*P*3$_2$21)** | ICSD Code 70054 |  |
| *a* (Å) | 4.145 (2) | 4.14883 (7) |
| *c* (Å) | 9.496 (2) | 9.4970 (4) |
| Crystallite size (nm) |  | 16.91 (10) |
| *R*$_{wp}$ (%) |  | 7.43 |

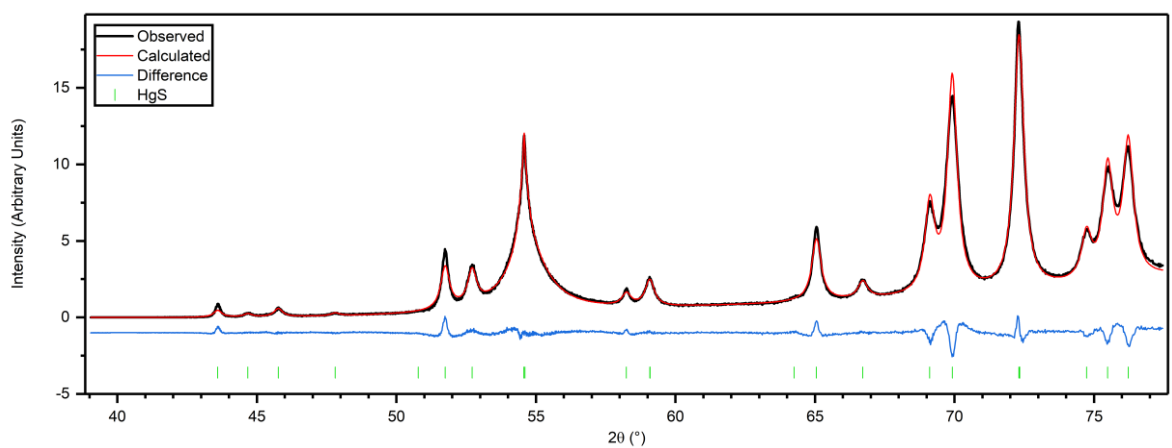

**Fig. S22.** Pawley fit to the Vermilion diffraction pattern. The difference plot is shown offset on the vertical scale. The broad peak centred on 2θ = 54.8° is due to diffraction by the graphitic vacuum window, and was fitted by inclusion of two pseudo-Voigt functions.

HgS (cinnabar) can be fitted to the Vermilion diffraction pattern (Fig. S22). Analysis of the significant peak broadening suggests a crystallite size of 16.91 (10) nm (Table S14).

**Minium (Rublev Colours)**

Table S15. Refined crystallographic parameters for the identified phase in Minium.

|  | Literature | Pawley Fit |
|---|---|---|
| **Pb$_3$O$_4$ (*P*4$_2$/*mbc*)** | ICSD Code 4106 |  |
| *a* (Å) | 8.811[a] | 8.81596 (3) |
| *c* (Å) | 6.563[a] | 6.56641 (2) |
| *R*$_{wp}$ (%) |  | 5.59 |

[a]Error estimates not reported.

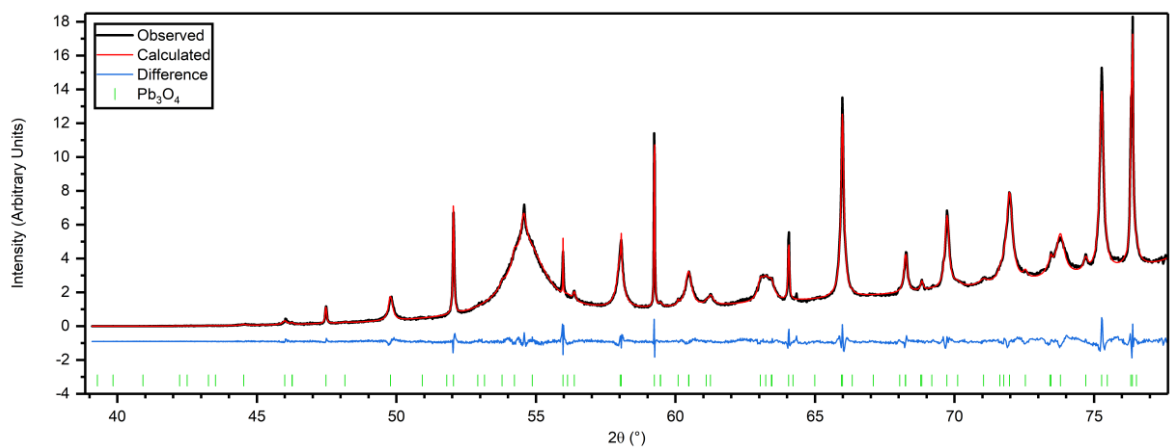

**Fig. S23.** Pawley fit to the Minium diffraction pattern. The difference plot is shown offset on the vertical scale. The broad peak centred on $2\theta = 54.8°$ is due to diffraction by the graphitic vacuum window, and was fitted by inclusion of two pseudo-Voigt functions.

All the peaks of the Minium paint diffraction pattern can be attributed to a single $Pb_3O_4$ phase (which has the mineral name minium). The pattern exhibits extremely anisotropic peak broadening (Fig. S23). An anisotropic crystallite size model [8] did not satisfactorily fit the pattern, with the largest improvement given by the Stephens anisotropic strain model [9]. For comparison, the $R_{wp}$ values for an isotropic peak shape, an anisotropic crystallite size model and an anisotropic strain model were 12.4%, 9.82% and 5.59% (Table S15) respectively. This result is consistent with previous studies [10] and it appears that microstrain anisotropy is an inherent property of the crystal structure.

### French Yellow Ochre (Michael Harding No. 133)

**Table S16.** Refined crystallographic parameters for the identified phases in French Yellow Ochre.

|  | Literature | Pawley Fit |
|---|---|---|
| **α-FeOOH (goethite) (*Pbnm*)** | ICSD Code 239324 | |
| $a$ (Å) | 4.6145 (8) | 4.5977 (4) |
| $b$ (Å) | 9.9553 (17) | 9.9215 (11) |
| $c$ (Å) | 3.0177 (5) | 3.0087 (4) |
| Crystallite size (nm) |  | 17.9 (4) |
| **$SiO_2$ (quartz) (*P3$_2$21*)** | ICSD Code 34644 | |
| $a$ (Å) | 4.9138 (2) | 4.91511 (2) |
| $c$ (Å) | 5.4052 (2) | 5.40583 (3) |
| Crystallite size (nm) |  | 126 (2) |
| **$TiO_2$ (rutile) (*P4$_2$/mnm*)** | ICSD Code 9161 | |
| $a$ (Å) | 4.5941 (1) | 4.5939 (2) |
| $c$ (Å) | 2.9589 (1) | 2.9589 (3) |
| Crystallite size (nm) |  | 141 (23) |
| $R_{wp}$ (%) |  | 9.76 |

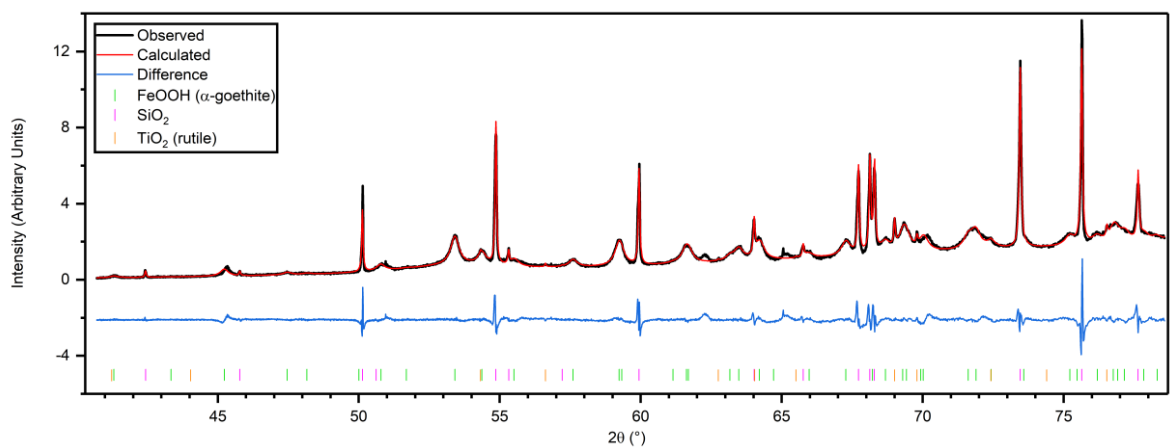

**Fig. S24.** Pawley fit to the French Yellow Ochre diffraction pattern. The difference plot is shown offset on the vertical scale.

A Pawley fit to the diffraction pattern of French Yellow Ochre (Fig. S24) based on α-FeOOH (goethite), $SiO_2$ (quartz) and $TiO_2$ (rutile) accounts for the majority of peaks, with weak unidentified peaks at $2\theta$ = 62.3° and 65.2°. Goethite was found to have small crystallites of mean diameter 17.9 (4) nm, whilst the quartz and rutile phases have much sharper peaks (Table S16). These two phases also show extremely poorly powder averaging and consequently no Rietveld fit was attempted.

**Raw Sienna (Michael Harding No. 120)**

**Table S17.** Refined crystallographic parameters for the identified phases in Raw Sienna.

|  | Literature | Pawley Fit |
|---|---|---|
| **α-FeOOH (goethite) (*Pbnm*)** | ICSD Code 239324 | |
| *a* (Å) | 4.6145 (8) | 4.6094 (4) |
| *b* (Å) | 9.9553 (17) | 9.9639 (11) |
| *c* (Å) | 3.0177 (5) | 3.0207 (2) |
| Crystallite size (nm) |  | 13.3 (2) |
| **SiO$_2$ (quartz) (*P3$_2$21*)** | ICSD Code 34644 | |
| *a* (Å) | 4.9138 (2) | 4.9144 (2) |
| *c* (Å) | 5.4052 (2) | 5.4050 (15) |
| Crystallite size (nm) |  | 130 (20) |
| **TiO$_2$ (rutile) (*P4$_2$/mnm*)** | ICSD Code 9161 | |
| *a* (Å) | 4.5941 (1) | 4.5872 (2) |
| *c* (Å) | 2.9589 (1) | 2.95790 (7) |
| Crystallite size (nm) |  | 220 (30) |
| *R$_{wp}$* (%) |  | 5.68 |

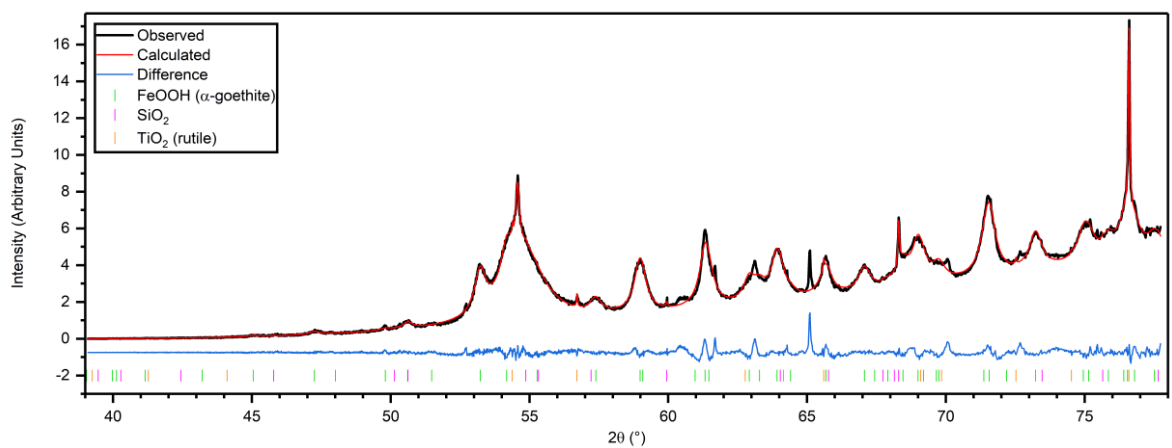

**Fig. S25.** Pawley fit to the Raw Sienna diffraction pattern. The difference plot is shown offset on the vertical scale. The broad peak centred on $2\theta = 54.8°$ is due to diffraction by the graphitic vacuum window, and was fitted by inclusion of two pseudo-Voigt functions.

Raw Sienna is composed mostly of goethite, with several peaks attributed to quartz and rutile though several expected peaks for these phases have no appreciable intensity due to poor powder-averaging (Fig. S25). The goethite mean crystallite size is 13.3 (2) nm (Table S17), similar to the size found in French Yellow Ochre (and Raw Umber, below). Several diffraction peaks remain unidentified.

**Raw Umber (Michael Harding No. 121)**

Table S18. Refined crystallographic parameters for the identified phases in Raw Umber.

|  | Literature | Pawley Fit |
|---|---|---|
| **α-FeOOH (goethite) (*Pbnm*)** | ICSD Code 239324 | |
| $a$ (Å) | 4.6145 (8) | 4.6016 (12) |
| $b$ (Å) | 9.9553 (17) | 9.969 (2) |
| $c$ (Å) | 3.0177 (5) | 3.0105 (4) |
| Crystallite size (nm) |  | 6.51 (13) |
| **SiO$_2$ (quartz) (*P3$_2$21*)** | ICSD Code 34644 | |
| $a$ (Å) | 4.9138 (2) | 4.91479 (12) |
| $c$ (Å) | 5.4052 (2) | 5.40545 (15) |
| Crystallite size (nm) |  | 410 (60) |
| $R_{wp}$ (%) |  | 4.06 |

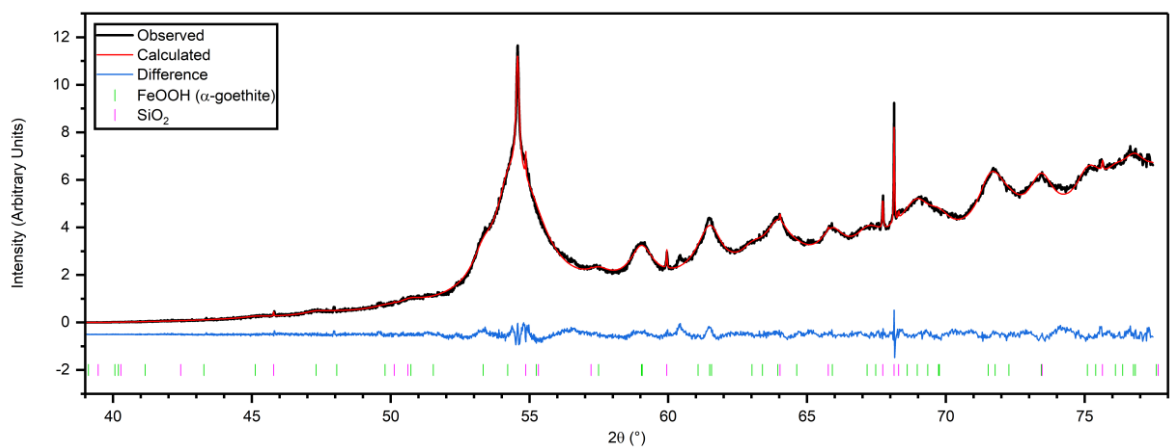

**Figure S26.** Pawley fit to the Raw Umber diffraction pattern. The difference plot is shown offset on the vertical scale. The broad peak centred on $2\theta = 54.8°$ is due to diffraction by the graphitic vacuum window, and was fitted by inclusion of two pseudo-Voigt functions.

As for French Yellow Ochre and Raw Sienna, Raw Umber consists mainly of goethite (Fig. S26) and poorly powder-averaged quartz, though no rutile peaks were observed in this case. The goethite peaks were exceptionally broad, yielding a crystallite size of 6.51 (13) nm (Table S18).

**Transparent Red Oxide (Michael Harding No. 220)**

**Table S19.** Refined crystallographic parameters for the identified phase in Transparent Red Oxide.

|  | Literature | Pawley Fit |
|---|---|---|
| **$Fe_2O_3$ (hematite) ($R\bar{3}c$)** | ICSD Code 7797 | |
| $a$ (Å) | 5.0324 (9) | 5.03661 (3) |
| $c$ (Å) | 13.7643 (4) | 13.7606 (4) |
| $L_x$ (nm) |  | 8.25 (2) |
| $L_z$ (nm) |  | 4.36 (2) |
| $R_{wp}$ (%) |  | 3.13 |

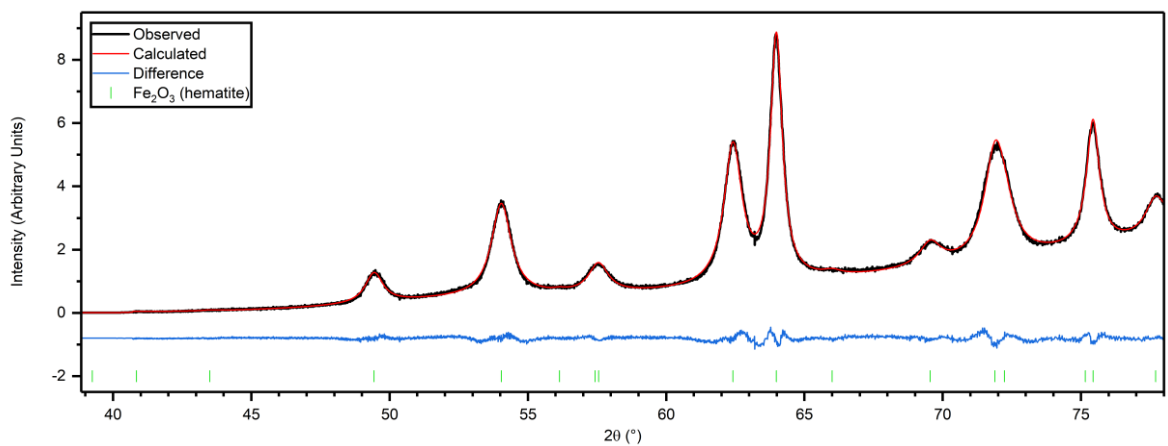

**Fig. S27.** Pawley fit to the Transparent Red Oxide diffraction pattern. The difference plot is shown offset on the vertical scale.

Transparent red oxide consists of $Fe_2O_3$ (hematite) (Fig. S27). The manufacturer notes the pigment's exceptionally small crystallite size and the diffraction peaks are broadened anisotropically. Application of the crystallite shape model of Ectors *et al* [8] suggests platy crystallites with an $L_x : L_z$ ratio ~2 assuming cylindrical crystallites (Table S19). It should be noted however that a fit using the

Stephens anisotropic strain model [9] instead of crystallite size is of comparable quality ($R_{wp}$ = 2.97%), albeit with several additional refined parameters.

**Cremnitz White (Michael Harding No. 307)**

Table S20. Refined crystallographic parameters for the identified phases in Cremnitz White.

|  | Literature | Pawley Fit | Rietveld Fit[a] |
|---|---|---|---|
| **2PbCO$_3$•Pb(OH)$_2$ ($R\bar{3}m$)** | Siidra et al. [11] | | |
| $a$ (Å) | 5.2475 (1) | 5.24372 (6) | 5.2442 (2) |
| $c$ (Å) | 23.6795 (7) | 23.68405 (10) | 23.6841 (3) |
| $L_x$ (nm) | | 230 (30) | 76 (8) |
| $L_z$ (nm) | | 32.8 (4) | 36.6 (10) |
| Weight fraction (%) | | | 81.0 (8) |
| **PbCO$_3$ (*Pmcn*)** | Antao & Hassan [12] | | |
| $a$ (Å) | 5.18324 (2) | 5.18265 (5) | 5.1820 (3) |
| $b$ (Å) | 8.49920 (3) | 8.50094 (9) | 8.5022 (4) |
| $c$ (Å) | 6.14746 (3) | 6.14231 (8) | 6.1424 (3) |
| Crystallite size (nm) | | 89 (2) | 72 (3) |
| Weight fraction (%) | | | 19.0[b] (8) |
| $R_{wp}$ (%) | | 7.2 | 21.0 |

[a]Preferred orientation of the hydrocerussite phase was fitted by an eight-term spherical harmonics model.
[b]This phase has poor powder averaging.

The analysis of this paint is described in more detail in the main article.

**Flake White (Michael Harding No. 703)**

Table S21. Refined crystallographic parameters for the identified phases in Flake White.

|  | Literature | Pawley Fit | Rietveld Fit |
|---|---|---|---|
| **2PbCO$_3$•Pb(OH)$_2$ ($R\bar{3}m$)** | Siidra et al. [11] | | |
| $a$ (Å) | 5.2475 (1) | 5.24405 (2) | 5.24403 (5) |
| $c$ (Å) | 23.6795 (7) | 23.6819 (2) | 23.6822 (4) |
| $L_x$ (nm) | | 136 (6) | 104 (6) |
| $L_z$ (nm) | | 32.6 (6) | 37.0 (14) |
| $r$[a] | | | 0.709 (6) |
| Weight fraction (%) | | | 56.2 (10) |
| **PbCO$_3$ (*Pmcn*)** | Antao & Hassan [12] | | |
| $a$ (Å) | 5.18324 (2) | 5.18236 (13) | 5.1834 (4) |
| $b$ (Å) | 8.49920 (3) | 8.4968 (2) | 8.4977 (6) |
| $c$ (Å) | 6.14746 (3) | 6.1435 (2) | 6.1410 (5) |
| Crystallite size (nm) | | 95 (2) | 71 (5) |
| Weight fraction (%) | | | 10.9[b] (5) |
| **ZnO ($P6_3mc$)** | NIST [13] | | |
| $a$ (Å) | 3.24983 (8) | 3.249919 (9) | 3.24992 (2) |
| $c$ (Å) | 5.2068 (1) | 5.20682 (3) | 5.20688 (7) |
| Crystallite size (nm) | | 268 (5) | 275 (11) |
| Weight fraction (%) | | | 32.9 (9) |
| $R_{wp}$ (%) | | 4.69 | 12.1 |

[a]March-Dollase factor.
[b]This phase has poor powder averaging.

The analysis of this paint is described in more detail in the main article.

## Zinc White (Michael Harding No. 103)

**Table S22.** Refined crystallographic parameters for the identified phase in Zinc White.

|  | Literature | Pawley Fit |
|---|---|---|
| **ZnO ($P6_3mc$)** | NIST [13] |  |
| $a$ (Å) | 3.24983 (8) | 3.249957 (6) |
| $c$ (Å) | 5.2068 (1) | 5.20654 (2) |
| Crystallite size (nm) |  | 251 (2) |
| $R_{wp}$ (%) |  | 9.01 |

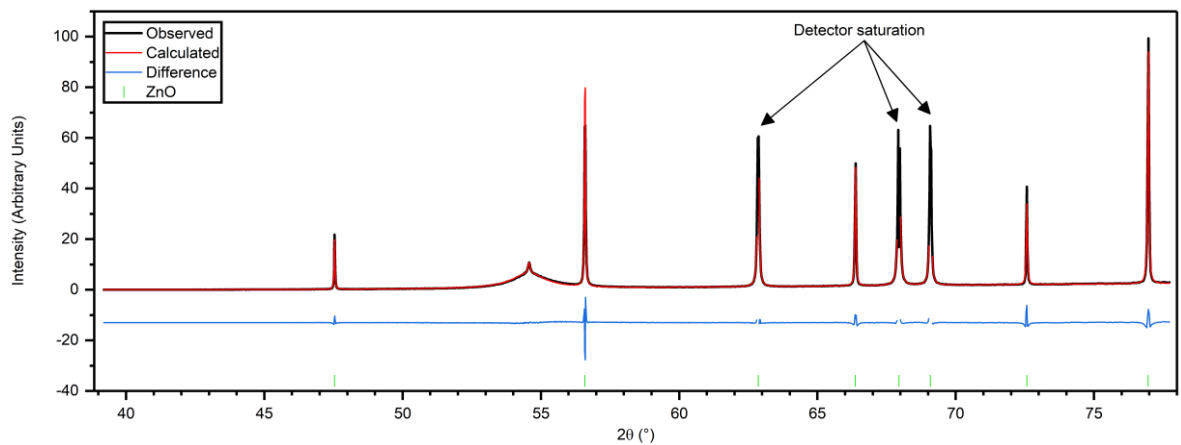

**Fig. S28**. Pawley fit to the Zinc White diffraction pattern. The difference plot is shown offset on the vertical scale. The broad peak centred on $2\theta = 54.8°$ is due to diffraction by the graphitic vacuum window, and was fitted by inclusion of two pseudo-Voigt functions.

ZnO (zincite) was the only crystalline phase observed in Zinc White (Table S22). Several peaks in the diffraction pattern of Zinc White caused detector saturation (Fig. S28), and the affected data points were excluded from the Pawley fit.

## Flemish White (Rublev Colours)

**Table S23.** Refined crystallographic parameters for the identified phase in Flemish White.

|  | Literature | Pawley Fit | Rietveld Fit |
|---|---|---|---|
| **(PbO)$_3$PbSO$_4$·H$_2$O ($P\bar{1}$)** | Steele et al [14] |  |  |
| $a$ (Å) | 6.3682 (2) | 6.3737 (5) | 6.3755 (5) |
| $b$ (Å) | 7.4539 (3) | 7.4520 (7) | 7.4487 (6) |
| $c$ (Å) | 10.2971 (4) | 10.2987 (6) | 10.2970 (7) |
| $\alpha$ (°) | 75.33 (3) | 75.351 (7) | 75.324 (6) |
| $\beta$ (°) | 79.40 (1) | 79.413 (7) | 79.416 (7) |
| $\gamma$ (°) | 88.34 (1) | 88.228 (8) | 88.267 (7) |
| Crystallite size (nm) |  | 42.5 (16) | 39.9 (13) |
| $R_{wp}$ (%) |  | 3.63 | 12.0 |

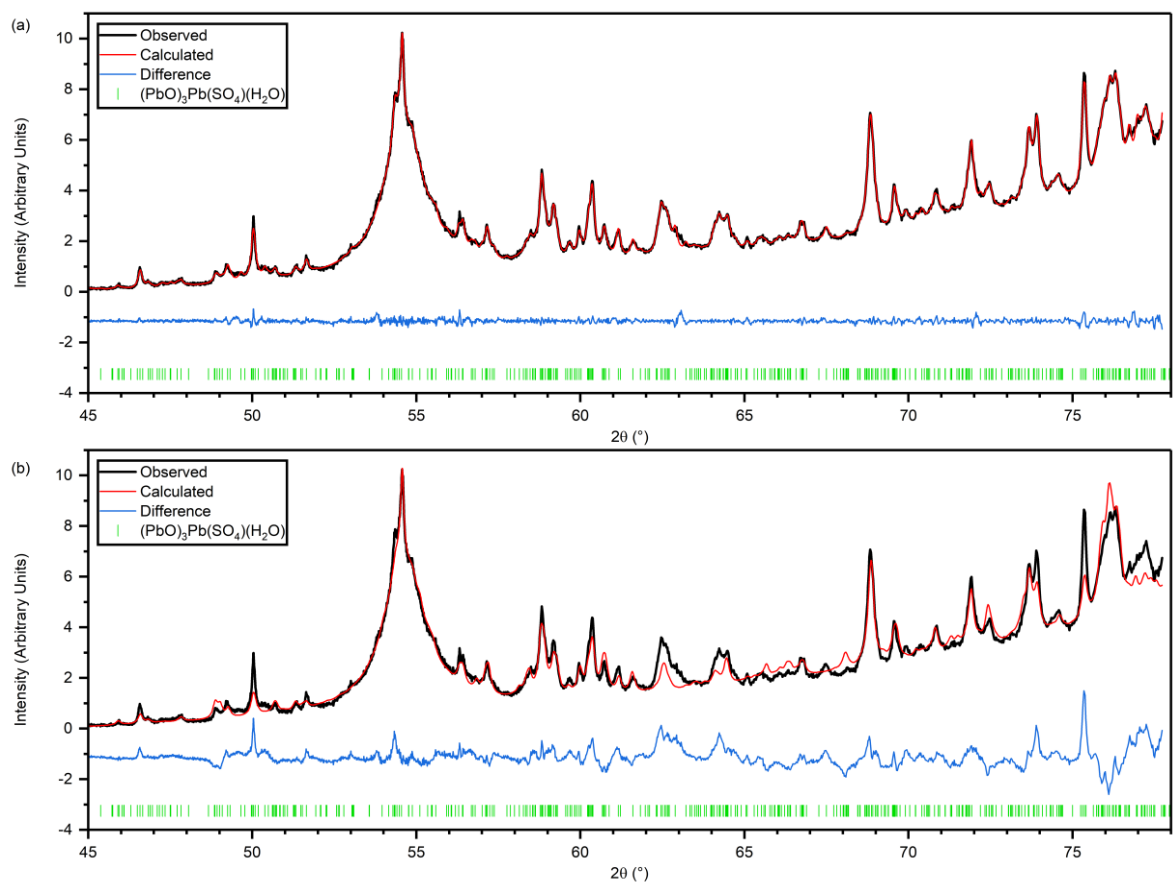

**Fig. S29.** (a) Pawley and (b) Rietveld fits to the Flemish White diffraction pattern. In each case the difference plot is shown offset on the vertical scale. The broad peak centred on $2\theta = 54.8°$ is due to diffraction by the graphitic vacuum window, and was fitted by inclusion of two pseudo-Voigt functions.

A Pawley fit to Flemish White based on $(PbO)_3PbSO_4 \cdot H_2O$ yields a good fit (Fig. S29a). However, the high density of allowed reflections means that a good agreement from a Pawley fit does not necessarily guarantee the correct structure has been assigned. To test whether the phase assignment is correct, a Rietveld refinement was carried out (Fig. S29b) in which the unit cell parameters were allowed to vary but the structural model was fixed to that of Steele *et al* [14]. The relative peak intensities are reproduced sufficiently well to provide convincing confirmation of the assignment.

**Titanium White No. 3 (Michael Harding No. 130)**

**Table S24.** Refined crystallographic parameters for the identified phases in Titanium White No. 3.

|  | Literature | Pawley Fit | Rietveld Fit |
|---|---|---|---|
| **TiO$_2$ (rutile) ($P4_2/mnm$)** | ICSD Code 9161 | | |
| $a$ (Å) | 4.5941 (1) | 4.593985 (10) | 4.59391 (3) |
| $c$ (Å) | 2.9589 (1) | 2.958962 (10) | 2.95900 (2) |
| Crystallite size (nm) | | 130.6 (6) | 122.9 (13) |
| Weight fraction (%) | | | 71.5 (6) |
| **ZnO ($P6_3mc$)** | NIST [13] | | |
| $a$ (Å) | 3.24983 (8) | 3.249915 (6) | 3.249855 (14) |
| $c$ (Å) | 5.2068 (1) | 5.206793 (16) | 5.20688 (3) |
| Crystallite size (nm) | | 192.6 (12) | 188 (3) |
| Weight fraction (%) | | | 28.5 (6) |
| $R_{wp}$ (%) | | 6.49 | 14.6 |

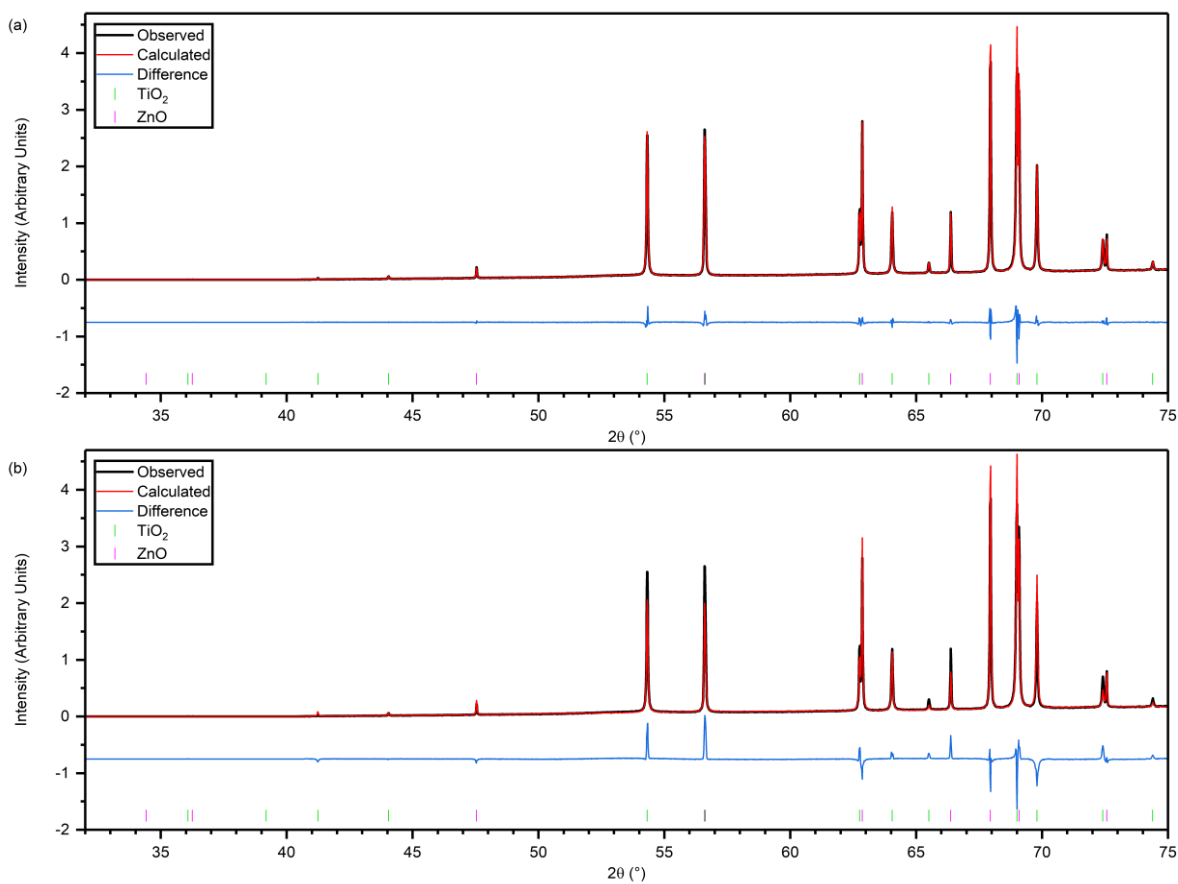

**Fig. S30.** (a) Pawley and (b) Rietveld fits to the Titanium White No. 3 diffraction pattern. In each case the difference plot is shown offset on the vertical scale.

Titanium White No. 3 paint was found to contain both $TiO_2$ (rutile) and ZnO (Fig. S30) in approximately a 2.5 : 1 ratio (Table S24).

**Ivory Black (Michael Harding No. 129)**

Table S25. Refined crystallographic parameters for the identified phases in Ivory Black.

|  | Literature | Pawley Fit | Rietveld Fit |
|---|---|---|---|
| **$Ca_5(PO_4)_3OH$ ($P6_3/m$)** | ICSD Code 56307 | | |
| $a$ (Å) | 9.4249 (4) | 9.4244 (5) | 9.4236 (4) |
| $c$ (Å) | 6.8838 (4) | 6.8848 (4) | 6.8850 (4) |
| Crystallite size (nm) | | 18.7 (12) | 21.9 (12) |
| Weight fraction (%) | | | 81.2 (13) |
| **$CaCO_3$ ($R\bar{3}c$)** | ICSD Code 18166 | | |
| $a$ (Å) | 4.9900 (2) | 4.9833 (5) | 4.9835 (7) |
| $c$ (Å) | 17.002 (1) | 17.059 (2) | 17.059 (3) |
| Crystallite size (nm) | | 29 (2) | 22.1 (13) |
| Weight fraction (%) | | | 18.8 (13) |
| $R_{wp}$ (%) | | 3.99 | 5.44 |

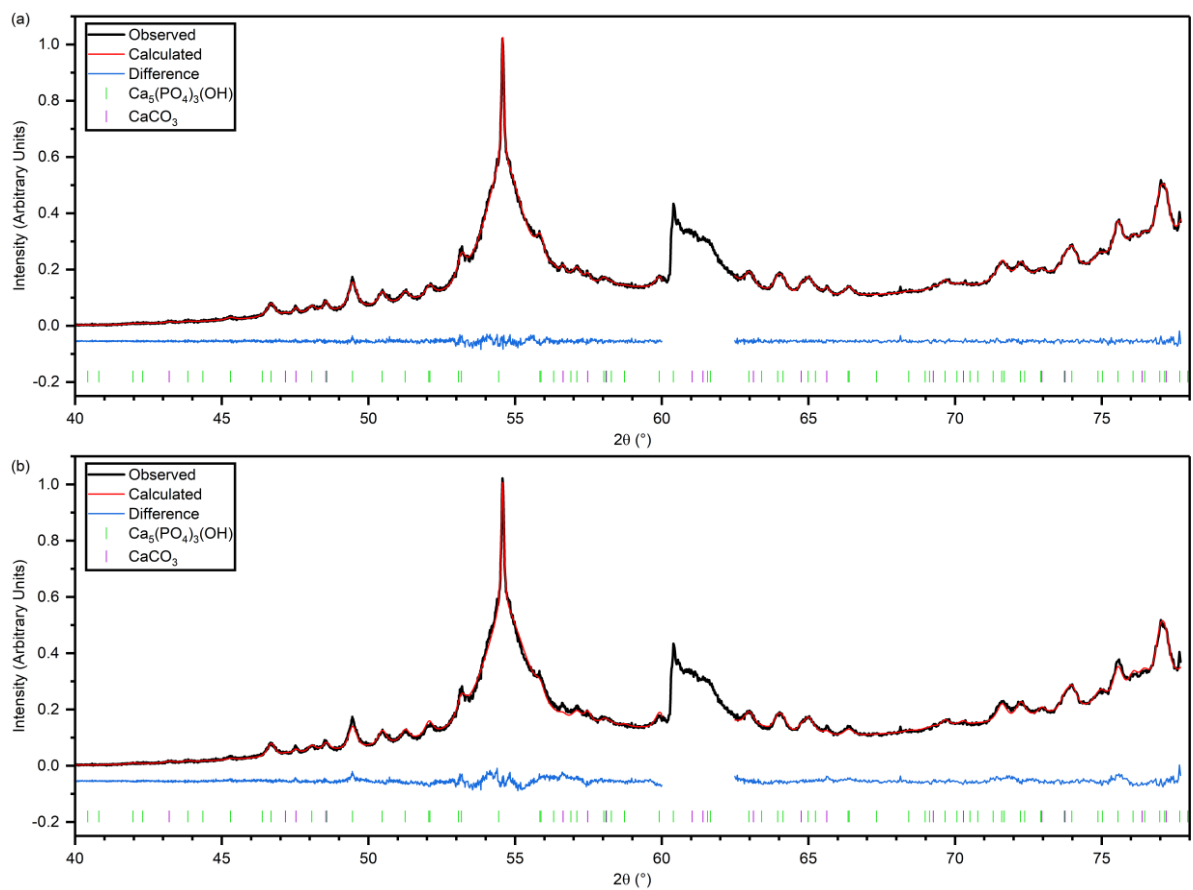

**Fig. S31.** (a) Pawley and (b) Rietveld fits to the Ivory Black diffraction pattern. In each case the difference plot is shown offset on the vertical scale. The broad peak centred on $2\theta = 54.8°$ is due to diffraction by the graphitic vacuum window, and was fitted by inclusion of two pseudo-Voigt functions.

The diffraction pattern of Ivory Black can be fitted with $Ca_5(PO_4)_3OH$ (hydroxyapatite) and $CaCO_3$ (calcite) (Table S25). A strong Ca K-edge is seen at $2\theta = 60.3°$ (Fig. S31). The edge and XAFS region were excluded from the fits. The black colour of the paint is expected to be due to carbonaceous material that is probably amorphous and is therefore undetected in this experiment. Strictly speaking, this paint does not meet the group A definition – that the pigment itself has been detected using XRD. However, the pigment is derived from burnt bone (according to the supplier) and the hydroxyapatite phase, and probably the calcite also, in therefore intimately linked with the carbonaceous material.

## 6.2    Group B Paints

**Manganese Blue (BLOCKX)**

**Table S26.** Refined crystallographic parameters for the identified phase in Manganese Blue.

|  | Literature | Pawley Fit |
|---|---|---|
| **ZnO ($P6_3mc$)** | NIST [13] |  |
| $a$ (Å) | 3.24983 (8) | 3.249900 (3) |
| $c$ (Å) | 5.2068 (1) | 5.206880 (9) |
| Crystallite size (nm) |  | 150.7 (4) |
| $R_{wp}$ (%) |  | 7.08 |

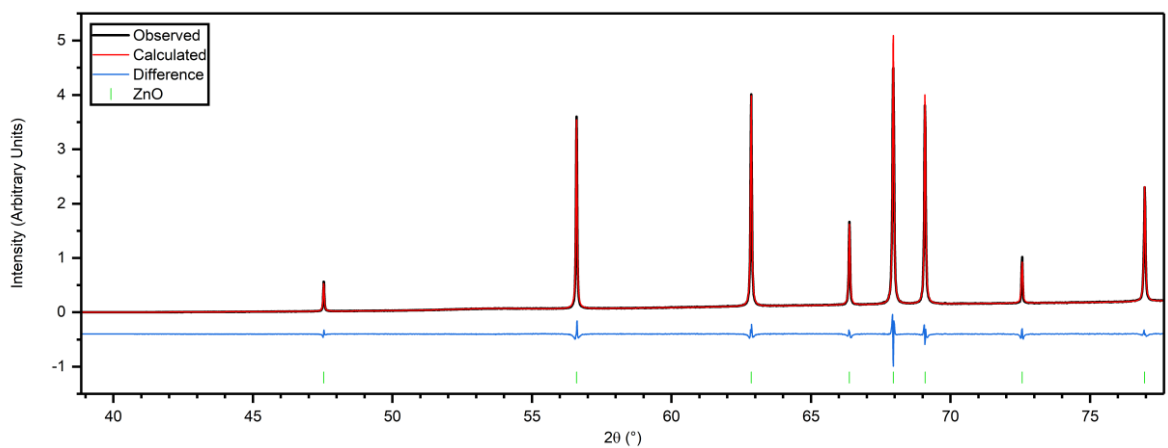

**Fig. S32**. Pawley fit to the Manganese Blue diffraction pattern. The difference plot is shown offset on the vertical scale.

ZnO was the only crystalline phase observed in Manganese Blue. The pigment in this paint is specified as phthalocyanine blue by the supplier.

### Dioxazine Violet (BLOCKX)

**Table S27.** Refined crystallographic parameters for the identified phase in Dioxazine Violet.

|  | Literature | Pawley Fit |
|---|---|---|
| **CaCO$_3$ ($R\bar{3}c$)** | ICSD Code 18166 | |
| $a$ (Å) | 4.9900 (2) | 4.98237 (8) |
| $c$ (Å) | 17.002 (1) | 17.0237 (3) |
| Crystallite size (nm) | | 53.8 (9) |
| $R_{wp}$ (%) | | 3.80 |

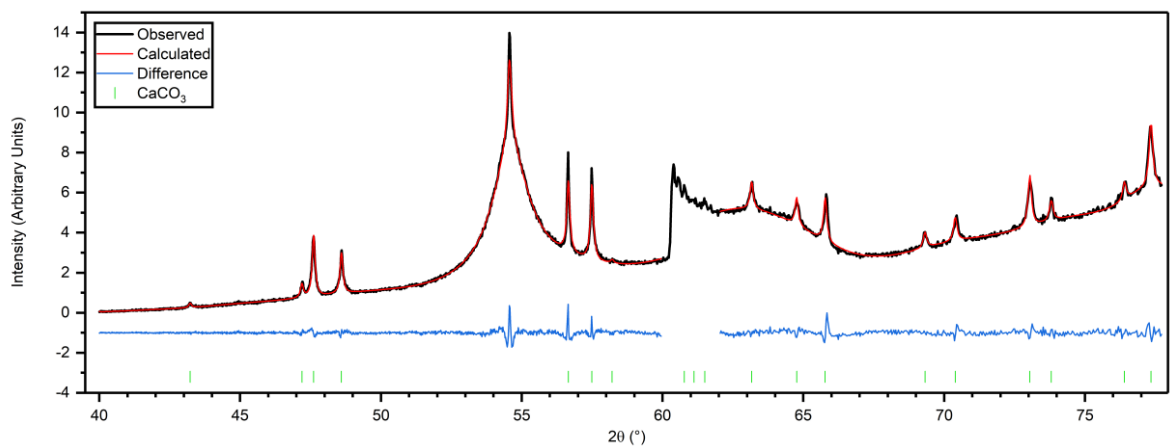

**Fig. S33**. Pawley fit to the Dioxazine Violet diffraction pattern. The difference plot is shown offset on the vertical scale. The broad peak centred on $2\theta = 54.8°$ is due to diffraction by the graphitic vacuum window, and was fitted by inclusion of two pseudo-Voigt functions.

CaCO$_3$ (calcite) was the only crystalline phase observed in Dioxazine Violet. A strong Ca K-edge is seen at $2\theta = 60.3°$ (Fig. S33). The edge and XAFS region were excluded from the Pawley fit. The pigment in this paint is specified as dioxazine by the supplier.

## Bright Yellow Lake (Michael Harding No. 109)

**Table S28.** Refined crystallographic parameters for the identified phase in Bright Yellow Lake.

|  | Literature | Pawley Fit |
|---|---|---|
| **TiO$_2$ (rutile) ($P4_2/mnm$)** | ICSD Code 9161 | |
| $a$ (Å) | 4.5941 (1) | 4.59370 (5) |
| $c$ (Å) | 2.9589 (1) | 2.95854 (6) |
| Crystallite size (nm) | | 143 (4) |
| $R_{wp}$ (%) | | 3.51 |

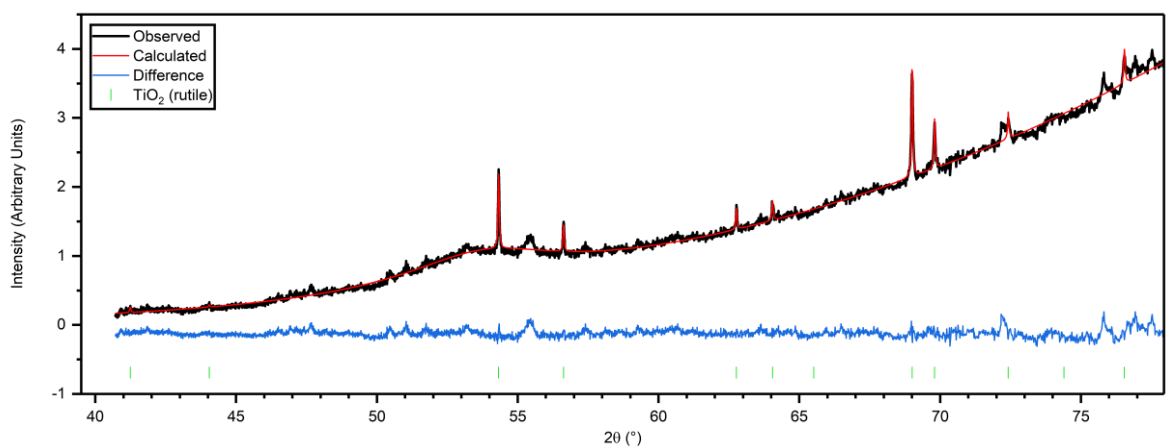

**Fig. S34**. Pawley fit to the Bright Yellow Lake diffraction pattern. The difference plot is shown offset on the vertical scale.

Rutile was the only phase identified in Bright Yellow Lake. However, several weak, relatively broad diffraction peaks remain unidentified. The pigment in this paint is specified as an arylamide (PY3) by the supplier.

## Pyrrolo-Vermilion (BLOCKX)

**Table S29.** Refined crystallographic parameters for the identified phases in Pyrrolo-Vermilion.

|  | Literature | Pawley Fit |
|---|---|---|
| **CaCO$_3$ ($R\bar{3}c$)** | ICSD Code 18166 | |
| $a$ (Å) | 4.9900 (2) | 4.98072 (5) |
| $c$ (Å) | 17.002 (1) | 17.02445 (9) |
| Crystallite size (nm) | | 76.0 (15) |
| **TiO$_2$ (rutile) ($P4_2/mnm$)** | ICSD Code 9161 | |
| $a$ (Å) | 4.5941 (1) | 4.59387 (2) |
| $c$ (Å) | 2.9589 (1) | 2.95881 (3) |
| Crystallite size (nm) | | 130 (2) |
| $R_{wp}$ (%) | | 9.67 |

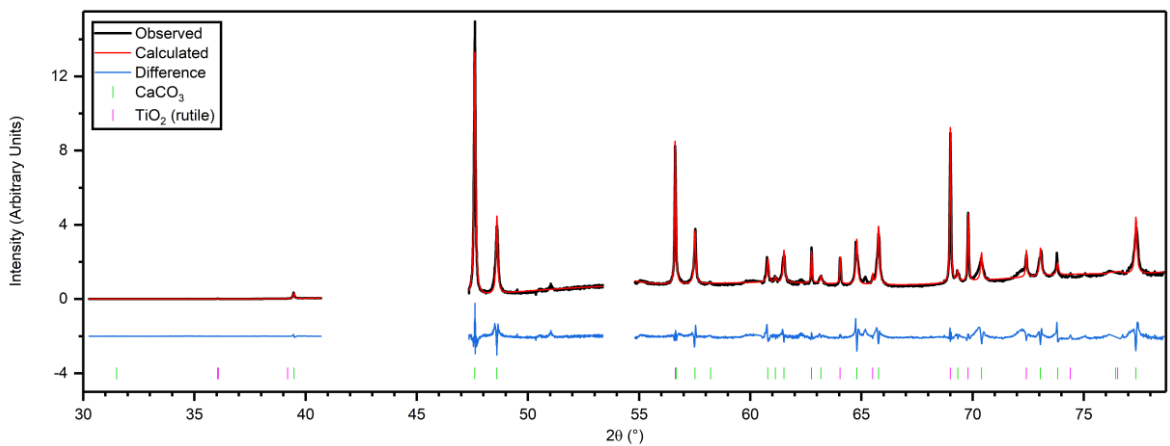

**Fig. S35**. Pawley fit to the Pyrrolo-Vermilion diffraction pattern. The difference plot is shown offset on the vertical scale.

Two segments of the data in the ranges $40.7° \leq 2\theta \leq 47.3°$ and $53.4° \leq 2\theta \leq 54.8°$ were corrupted at the data acquisition stage and were excluded from the fit (Fig. S35). The diffraction pattern shows that Pyrrolo-Vermilion contains $CaCO_3$ (calcite) and $TiO_2$ (rutile). As for Bright Yellow Lake, there are several weak unidentified diffraction peaks. The pigment in this paint is specified as pyrrolo orange (PO73) by the supplier.

## 6.3 Group C Paints

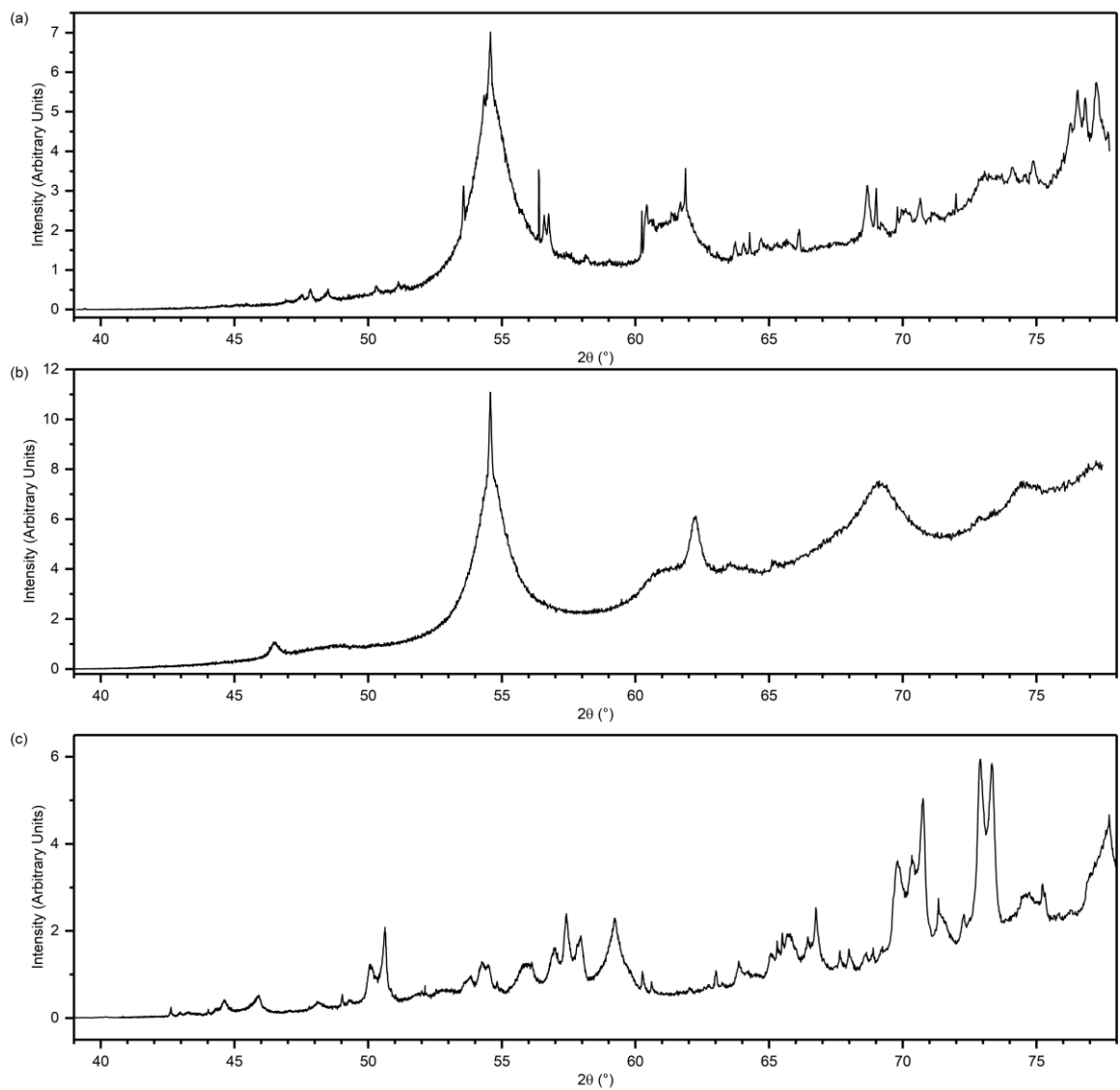

**Fig. S36.** Diffraction patterns of the Group C paints. (a) Terre Vert (Michael Harding No. 115), (b) Viridian (BLOCKX) and (c) Orange Molybdate (Rublev Colours). The broad peak centred on $2\theta = 54.8°$ in (a) and (b) is due to diffraction by the graphitic vacuum window. The Terre Vert pattern (a) has a Ca-K absorption edge at $2\theta = 60.3°$.

Although the Group C paints have obvious diffraction peaks (Fig. S36), no phases have been identified. Terre Vert contains the 'green earth' pigment (PG23) and there are therefore a relatively large number of candidate phases, and there may also be a complex mixture present. Orange Molybdate is expected to contain lead chromate molybdate (PR104). However, attempts to fit the pattern with the appropriate phase candidates were unsuccessful. The Viridian pattern shows a small number of broad, though variable-width, diffraction peaks. The peak positions do not correspond to those predicted for a range of plausible candidate phases. As noted in Section 3 of the main article, the pigment is expected to be $Cr_2O_3 \cdot 2H_2O$ but there are no published crystal structures for this phase. It seems likely that the pattern recorded in this work contains structural information but it is not interpretable without, for example, reference to structural models that attempt to explain the lack of long-range crystallinity.

## 6.4 Group D Paints

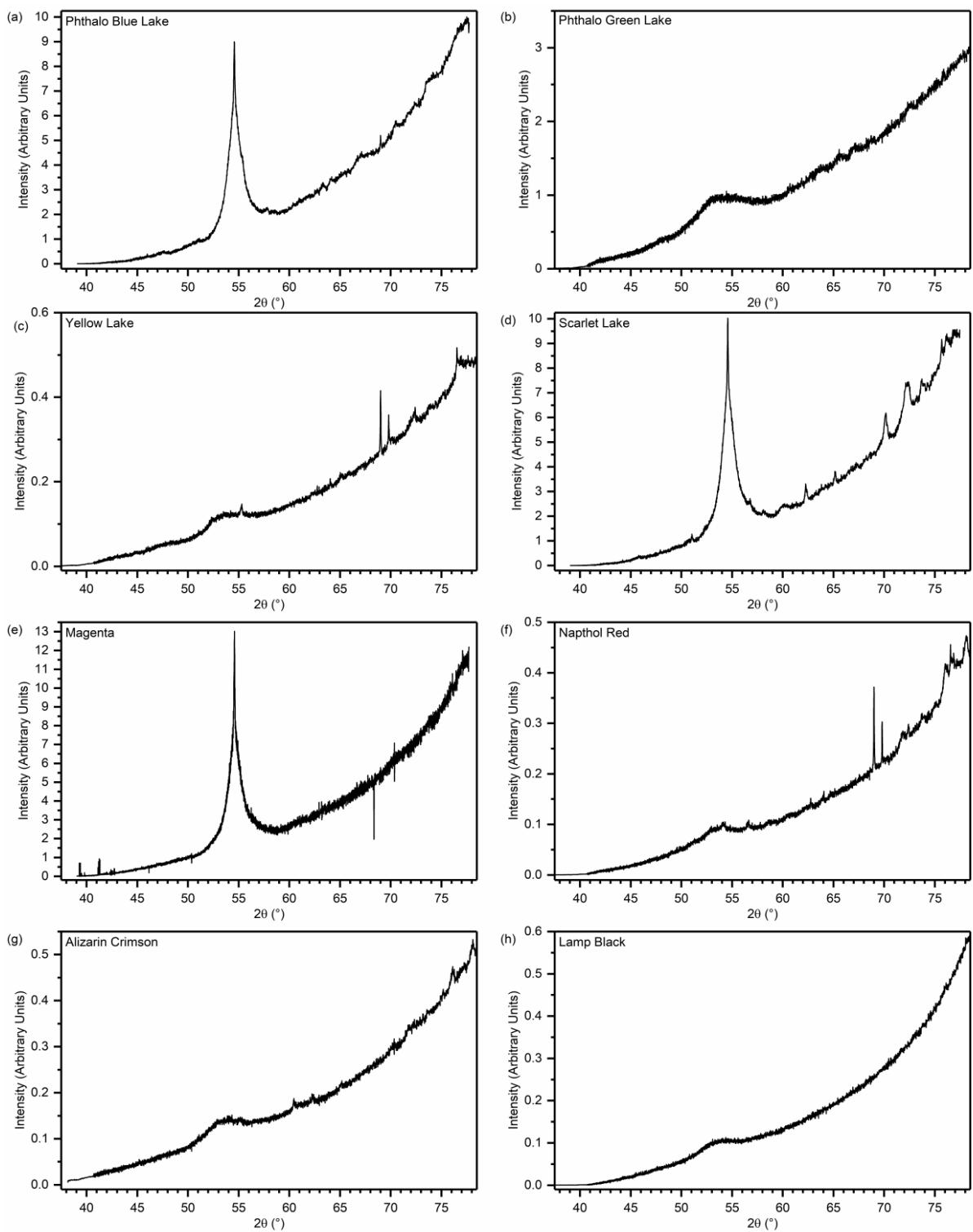

**Fig. S37.** Diffraction patterns of the Group D paints. (a) Phthalo Blue Lake (Michael Harding No. 209), (b) Phthalo Green Lake (Michael Harding No. 213), (c) Yellow Lake (Michael Harding No. 110), (d) Scarlet Lake (Michael Harding No. 205), (e) Magenta (Michael Harding No. 303), (f) Naphthol Red (Michael Harding No. 301), (g) Alizarin Crimson (Michael Harding No. 302), (h) Lamp Black (Michael Harding No. 128). (a), (d) and (e) contain a broad peak centred on $2\theta = 54.8°$ that is due to diffraction by the graphitic vacuum window. (c) and (f) contain diffraction peaks from $TiO_2$, rutile ($2\theta = 69.0°, 69.8°$) and a Ti K-edge at $2\theta = 76.2°$, believed to be due to 'break-through' of the white paint layer on the canvas beneath the paint samples.

Of the group D paints, only the diffraction patterns of Magenta and Lamp Black are truly devoid of any diffraction peaks (excluding the vacuum window peaks). The very sharp features in the Magenta pattern are noise peaks. It is assumed that the observed diffraction peaks, other than the rutile peaks, are due to the synthetic organic pigments that are expected to be present in these paints (except Lamp Black). For example, it is possible to do a Pawley fit of the Scarlet Lake pattern using a published crystal structure of naphthol (PR170). However, organic pigments typically have low crystal symmetry and relatively large unit cells leading to a very high density of diffraction peaks in the $2\theta$ range observed. Together with the low-quality observed diffraction patterns, it is not possible to uniquely identify the pigments present or to extract meaningful crystallographic parameters using an assumed crystal structure.